# Deformation and breakup of a ferrofluid compound droplet migrating in a microchannel under a magnetic field: A phase-field-based multiple-relaxation time lattice Boltzmann study

Parham Poureslami [a], Mohammad Majidi [b], Javad Ranjbar Kermani [a],

Mohamad Ali Bijarchi [a, *]

[a] Department of Mechanical Engineering, Sharif University of Technology, Tehran, Iran
[b] Department of Mechanical Engineering, College of Engineering, University of Tehran, Tehran, Iran

[*] Corresponding author: bijarch@sharif.edu

## Abstract

Though ubiquitous in many engineering applications, including drug delivery, the compound droplet hydrodynamics in confined geometries have been barely surveyed. For the first time, this study thoroughly investigates the hydrodynamics of a ferrofluid compound droplet (FCD) during its migration in a microchannel under the presence of a pressure-driven flow and a uniform external magnetic field (UEMF) to manipulate its morphology and retard its breakup. Finite difference and phase-field multiple-relaxation time lattice Boltzmann approaches are coupled to determine the magnetic field and ternary flow system, respectively. Firstly, the influence of the magnetic Bond number ($Bo_m$) on the FCD morphology is explored depending on whether the core or shell is ferrofluid when the UEMF is applied along $\alpha = 0°$ and $\alpha = 90°$ relative to the fluid flow. It is ascertained that imposing the UEMF at $\alpha = 0°$ when the shell is ferrofluid can postpone the breakup. Intriguingly, when the core is ferrofluid, strengthening the UEMF enlarges the shell deformation. Afterward, the effects of the Capillary number (Ca), density ratio, viscosity ratio, radius ratio, and surface tension coefficients are scrutinized on the FCD deformation and breakup. The results indicate that augmenting the core-to-shell viscosity and density ratios accelerates the breakup process. Additionally, surface tension between the core and shell suppresses the core deformation. Moreover, increasing the Ca number intensifies the viscous drag force exerted on the shell, flattening its rear side, which causes a triangular-like configuration. Ultimately, by varying $Bo_m$ and Ca numbers, five distinguished regimes are observed, whose regime map is established.

# 1 Introduction

Generated by encapsulating a liquid droplet (core) in another immiscible droplet (shell) [1], a compound droplet (double emulsion), has gained painstaking attention in the wake of its numerous applications, from which one may allude to pharmaceutics [2], drug delivery [3, 4], the food industry [5], agriculture, lab-on-a-chip [6], and cosmetic. Deformation and breakup are the regular phenomena a compound droplet experiences in the mentioned applications, inducing a change in their morphology and properties. Accordingly, not only does the fundamental survey of compound droplet deformation and breakup instigate the appreciation of their underlying physics, but it also extends their applications, which is the primary stimulant for performing this research.

Deformation and breakup of the compound droplet have been investigated in extensional flow [7], simple shear flow [8], microchannels [9], capillary tubes [10], and other specific configurations [11]. Regarding extensional flow, Qu and Wang [12] simulated the stability and deformation of a compound droplet in an extensional flow exploiting a 3D spectral boundary element. Exploring the impacts of the Capillary number (Ca), inner droplet size, surface tension ratio, and viscosity on the compound droplet stability, they asserted that the smaller the core, the more stable the compound droplet. Moreover, the initial location of the compound droplet can affect not only the droplet's stability but also its migration velocity. Santra et al. [13] surveyed the concurrent influences of an electric field and extensional flow on the deformation of a compound droplet. They posited that in the extensional flow, the inner and outer droplets merely deform into the oblate and prolate shapes, respectively, while different morphologies may happen when an electric field is brought to bear. Utilizing an electric current model and the phase-field method, Hao et al. [14] simulated the deformation of an electric-field-mediated compound droplet in an extensional flow. The results imply that the compound droplet deformation may be characterized by the electrophysical parameters of the fluid, such as the conductivity ratio, extensional flow strength, and electric field strength. Concerning the shear flow, examining the influences of the Capillary number, radius ratio, and viscosity ratio of outer to inner droplets, Chen et al. [15] numerically delved into the hydrodynamics of a compound droplet in shear flow. They observed four distinct breakup regimes and provided phase diagrams to elucidate the critical Capillary number at which the breakup occurs. Luo et al. [16] numerically explored the deformation of a compound droplet in a simple shear flow via a front-tracking method. They ascertained that the deformation of the compound droplet is smaller than that of a simple droplet, which is consistent with Hua et al. [17], who numerically concentrated on the deformation dynamics of a compound droplet in a shear flow. Liu et al. [18] studied the deformation and breakup of a compound droplet via a three-phase lattice Boltzmann method (LBM) in an oscillatory shear flow. They expressed that the maximum deformation parameter deviates from the linearity while Ca > 0.35. Furthermore, the critical Ca number determining the transition from deformation to breakup is pertinent to the confinement ratio. They also alleged that raising the oscillatory period curtails the critical Capillary number and alters the breakup mode. Generally speaking, although the compound droplet transport in a media encompassing constrictions has several authentic applications, including oil recovery and drug delivery [19], the number of studies allotted to the deformation of a compound droplet in a confined geometry (e.g., capillary tubes and microchannels) is inconsequential compared to those focused on the compound droplet deformation and breakup in the shear flow. Also, in light of the confinement effects emanating from the wall, the hydrodynamics of the compound droplets in shear and extensional flows cannot be generalized to microchannels and other microfluidic devices, necessitating further studies to illuminate their morphology in the microfluidic applications [20].

Given the surveys concentrating on the morphology of a compound droplet in a confined geometry, taking the effects of the Ca number and radius ratio into account, Chen et al. [21] experimentally examined the breakup of a double emulsion in the orifice of a tapered nozzle and constructed a phase diagram, revealing the behavior of a compound droplet under various circumstances. Similar to the previous research, Li et al. [22] explored the breakup of a double emulsion flowing through a tapered nozzle and contended that as the tilt angle of the nozzle surpasses 9°, the breakup is prevented. Ho and Vo [23] numerically scrutinized the deformation and breakup of a multi-core compound droplet in an axisymmetric T-junction via the front-tracking method. They reported three breakup regimes and determined the flow conditions under which the transition from non-breakup to breakup may take place. Exploiting the volume of fluid (VOF) method, Borthakur et al. [10] investigated the dynamics of a compound droplet migrating inside a capillary tube, presuming the axisymmetric condition. They concluded that the deformation of the core and shell relies on the Ca number; however, the influence of the Capillary number on the core deformation is more conspicuous. Sattari et al. [24] numerically studied the deformation of a compound droplet traversing inside a microcapillary tube in non-Newtonian and Newtonian ambient fluids. They maintained that whereas linearly boosting along the tube axis in a Newtonian ambient, the center of mass of inner and outer droplets' position indicates a non-linear behavior versus time in a non-Newtonian ambient. Additionally, the properties of the ambient fluid strikingly affect the compound droplet characteristics, including average acceleration, average velocity, and eccentricity. In view of its momentousness in microfluidic devices and micro-electro-mechanical systems (MEMS), the hydrodynamics of a compound droplet transporting in microchannel merit more attention, whereas investigations allocated to this topic are meager [9]. Simulating the multiphase flow by the volume of fluid method and capturing the interface features via level set methods, Che et al. [20] conducted research into the flow structure within a compound droplet moving inside a microchannel. They argued that though attenuating the vortices in the shell, raising the core size accentuates the vortices in the core. In addition, the greater the Capillary number, the more deformed the compound droplet becomes and the higher its eccentricity. By contrast, the viscosity ratio trivially influences the compound droplet shape. Nguyen et al. [25] surveyed the deformation and breakup of a compound droplet traveling in a microchannel, in the middle of which a cone was included. They expressed that depending on the inertia, capillary, and surface tension forces, the compound droplet may undergo deformation and breakup regimes, each of which was divided into two distinct modes. Scrutinizing the hydrodynamics of a compound droplet moving in a fully developed flow under the Stokes regime, Thammanna Gurumurthy and Pushpavanam [26] posited that compared to a single droplet, the existence of the core changes the compound droplet shapes and increases its deformation. However, in the mentioned studies, the authors did not apply an external magnetic field to control the morphology of a compound droplet and retard its breakup while migrating inside a channel, a crucial gap on which the current scholarship is focused.

Ferrofluid is a colloidal suspension constructed from nanoscale ferromagnetic particles in a base liquid [27], which is vastly utilized in different engineering facets, including microfluidics [28], heat transfer amelioration [29], drug delivery [30], lab-on-chip [31], and droplet manipulation [32], generation [33] and splitting [34]. The capability to be manipulated remotely by an external magnetic field is the appealing advantage of ferrofluids, based on which numerous investigations have been fulfilled to amend their hydrodynamics and heat transfer characteristics [35]. However, the morphology of the ferrofluid droplets under the concurrent influences of the magnetic field and hydrodynamic flows has not been unveiled yet. Concentrating on the breakup of a ferrofluid droplet in a shear flow under a uniform external magnetic field (UEMF), Hassan and Wang [36] studied the effects of viscosity ratio, magnetic field direction and strength on the breakup incidence at low Reynolds numbers. They contended that the magnetic field angles of 45° and 90° prompt the breakup even at low Ca numbers, while at the angles of 0° and 135°, the magnetic field impedes the breakup. Utilizing the phase-field simplified LBM, Zhang et al. [37] assessed the evolvement process of a ferrofluid droplet suspended between air and liquid layer when the magnetic field is applied vertically. They claimed that the elongated velocity, height of the center of mass, and aspect ratio of the ferrofluid droplet are pertinent to magnetic Bond number ($Bo_m$). Guilherme et al. [7] proposed a 3D computational investigation to discern the effect of a UEMF on the ferrofluid droplets in planar extensional flows. They demonstrated how UEMF intensity and direction can change the rheology of ferrofluid droplets. Abdo et al. [38] conducted numerical research into the single ferrofluid droplet subjected to the simultaneous impacts of small amplitude oscillatory shear and UEMF. Placing the ferrofluid droplet in a 3D domain suspended in a non-magnetizable fluid, they realized how UEMF orientation and shear oscillation frequency could affect the droplet shape.

Though applying a magnetic field to manipulate a ferrofluid compound droplet (FCD) has immediate applications, notably in targeted drug delivery, investigations devoted to this paramount issue are quite elusive [3, 39]. Huang et al. [3] investigated the formation and morphology of a double emulsion in a microfluidic system under a non-uniform magnetic field engendered by a permanent magnet. The results indicate that the lower the downstream distance of the magnet from the ferrofluid junction, the greater the core droplet aspect ratio. Majidi et al. [39] numerically scrutinized the deformation and breakup of a single-core FCD in a shear flow. Imposing a UEMF at various angles to acquire further control over the FCD, they assessed the influences of Ca and $Bo_m$ numbers on the compound droplet deformation when either the core or shell is ferrofluid. They appreciated that boosting the $Bo_m$ number augments the elongation of inner and outer droplets along the UEMF direction. To the best of the authors' knowledge, no scrutiny has been conducted in the literature to investigate the hydrodynamics of an FCD migrating in a fluid flow in confined geometries under the presence of a UEMF, necessitating the scholars to scrutinize the phenomenon so that the morphology and hydrodynamics of the FCD can be magnetically manipulated where required.

Contemplating the thorough literature survey, it is crucial to investigate the deformation and breakup of an FCD when transporting in a microchannel under the concurrent effects of the fluid flow and magnetic field to manipulate its morphology such that the breakup is postponed in targeted drug delivery application. Thus, for the first time, this study numerically surveys the hydrodynamics of an FCD migrating in a microchannel when a UEMF is imposed. The hybrid multiple-relaxation-time (MRT) lattice Boltzmann-finite difference technique is implemented to explore the multiphase flow system and the magnetic field. The influences of the $Bo_m$ number, Ca number, density ratio, viscosity ratio, radius ratio, and surface tension coefficients are examined on the morphology of the FCD when either the core or shell is ferrofluid at disparate magnetic field angles. Moreover, the regime map of the FCD breakup is established based on the non-dimensional parameters, including $Bo_m$ and Ca numbers. The research fills the gap between the fundamental subject of compound droplet hydrodynamics and site-specific drug delivery, which is an overriding challenge in bio-medical applications.

## 2 Numerical methodology and problem definition

Multiphase lattice Boltzmann models can be generally divided into four distinct categories, among which the phase-field model is an outstanding scheme for simulating the multiphase flows, particularly when the impact of the surface tension is pronounced [40]. Additionally, the phase-field model is a tailored scheme not only for parallel computing but also for complex boundary conditions [41]. Accordingly, in this paper, the phase-field LBM is adopted to describe the ternary fluid system, whose governing equations are elaborated on briefly in this section. Furthermore, the finite difference method (FDM) is utilized to discretize the magnetic field equations.

### 2.1  FDM for the magnetic field

The magnetic field can be procured by the Maxwell's equation for the non-conducting fluid:

$$\nabla \cdot \mathbf{B} = 0 \tag{1}$$

and

$$\nabla \times \mathbf{H} = 0 \tag{2}$$

where **H** is the magnetic field intensity, whereas **B** denotes the magnetic flux density, which in turn may be attained as follows:

$$\mathbf{B} = \mu_0 (1 + \chi)\mathbf{H} = \mu \mathbf{H} \tag{3}$$

where $\mu$, $\mu_0$, and $\chi$ represent the magnetic permeability, vacuum permeability, and the magnetic susceptibility, respectively. The value of $\mu_0$ is $4\pi \times 10^{-7}$ N/A². By presuming magnetization (**M**) as a linear function of the magnetic field, one may have:

$$\mathbf{M} = \chi \mathbf{H} \tag{4}$$

In addition, owning to the curl-free constraint, the magnetic field may be expressed as:

$$\mathbf{H} = -\nabla \psi \tag{5}$$

where $\psi$ stands for the magnetic scalar potential. By incorporating Eqs. (3) and (5) into (1), Maxwell's equation is reduced to:

$$\nabla \cdot [(1 + \chi)\nabla \psi] = 0 \tag{6}$$

Eq. (2) is satisfied as well. The second-order (central) isotropic FDM is exploited to obtain $\psi$ values across the domain. The reader is referred to the article of Kumar [42] for the detailed explanations regarding the isotropic FDM scheme. Subsequently, through Eq. (5), the magnetic field intensity is attained, after which the magnetic force (**F**$_\mathbf{m}$) can be calculated:

$$\mathbf{F_m} = -\frac{1}{2}\mu_0 |\mathbf{H}|^2 \nabla \chi \tag{7}$$

It is supposed that $\chi$ is the harmonic mean of magnetic susceptibilities of the core ($\chi_1$), shell ($\chi_2$), and ambient fluid ($\chi_3$):

$$\frac{1}{1+\chi} = \sum_{i=1}^{3} \frac{\phi_i}{1+\chi} \tag{8}$$

where $\phi_i$ is the phase-field variable of fluid *i*. Eventually, $\nabla \chi$ is achieved by:

$$\nabla \chi = \sum_{i=1}^{3} \frac{d\chi_i}{d\phi_i} \nabla \phi_i = -\frac{\sum_{i=1}^{3} \frac{\nabla \phi_i}{1+\chi_i}}{\left(\sum_{i=1}^{3} \frac{\phi_i}{1+\chi_i}\right)^2} \tag{9}$$

## 2.2 MRT-LBM for the flow field

The isothermal and incompressible multiphase flow considered in this research is governed by the continuity and Navier–Stokes equations, respectively, as follows [43]:

$$\nabla \cdot \mathbf{u} = 0 \tag{10}$$

and

$$\rho \left( \frac{\partial \mathbf{u}}{\partial t} + \mathbf{u} \cdot \nabla \mathbf{u} \right) = -\nabla p + \nabla \cdot (\mu [\nabla \mathbf{u} + (\nabla \mathbf{u})^T]) + \mathbf{F}_s + \mathbf{F}_b + \mathbf{F}_m \tag{11}$$

where $\rho$ is the density, $\mathbf{u}$ is the macroscopic velocity, and $p$ denotes the pressure. Also, $\mathbf{F}_s$ is the surface tension force, which is clarified in Ref. [44]. Furthermore, $\mathbf{F}_b = \rho g$ represents the gravitational force. Eventually, $\mathbf{F}_m$ stands for the magnetic force, which can be computed by Eq. (7).

Providing a more stable solution, notably when the viscosity and density ratio of two phases is augmented, the MRT-LBM is implemented to solve Eqs. (10) and (11) [45, 46]. The Two-dimensional nine-velocity (D2Q9 model) is exploited in this paper, whose structure is provided in Fig. S1 of the Supplementary Material. Herein, the MRT formulation, originally proposed by Zu and He [43] and modified by Fakhari et al. [47] is implemented:

$$f_\alpha(\mathbf{x} + \mathbf{e}_\alpha \delta t, t + \delta t) - f_\alpha(\mathbf{x}, t) = -\mathbf{M}^{-1} \mathbf{\Lambda} \mathbf{M} \left( f_\alpha(\mathbf{x}, t) - \bar{f}_\alpha^{eq}(\mathbf{x}, t) \right) + F_\alpha(\mathbf{x}, t) \tag{12}$$

where $\delta t$ is the time step, $\mathbf{e}_\alpha$ denotes the particle velocity in the α-direction, while $\mathbf{M}$ denotes the orthogonal matrix projecting the distribution functions onto the moment space [48]. Also, $\mathbf{\Lambda}$ is the diagonal relaxation matrix whose elements are selected as follows [49]:

$$\mathbf{\Lambda} = diag \left( 1,1,1,1,1,1, \frac{1}{1/2 + \tau}, \frac{1}{1/2 + \tau} \right) \tag{13}$$

where $\tau$ is the relaxation time of the flow field and can be described by [45]:

$$\tau = \frac{\nu}{c_s^2 \delta t} \tag{14}$$

where $\nu$ is the fluid kinematic viscosity, and $c_s$ is the lattice sound speed, whose value is $1/\sqrt{3}$. It is worth mentioning that in Eq. (12), $f_\alpha(\mathbf{x}, t)$ is the velocity-based distribution for an incompressible fluid, whose modified equilibrium may be described by [47]:

$$\bar{f}_\alpha^{eq} = f_\alpha^{eq} - \frac{1}{2} F_\alpha(\mathbf{x}, t) \tag{15}$$

Therefore, the equilibrium distribution is achieved by subtracting half of the forcing term from the regular equilibrium distribution function so that the collision step may be simplified. In Eq. (15), $f_\alpha^{eq}$ takes the following form:

$$f_\alpha^{eq} = p^* \omega_\alpha + (\Gamma_\alpha - \omega_\alpha) \tag{16}$$

where $p^* = p/\rho c_s^2$ is the normalized pressure, while the weight coefficient $\omega_\alpha$ for the D2Q9 model is given as follows [50, 51]:

$$\omega_\alpha = \begin{cases} \frac{4}{9}, & \alpha = 0 \\ \frac{1}{9}, & \alpha = 1-4 \\ \frac{1}{36}, & \alpha = 5-8 \end{cases} \quad (17)$$

In addition, $\Gamma_\alpha$ is the non-dimensional distribution function, which can be procured through the following relation [52]:

$$\Gamma_\alpha = \omega_\alpha \left[ 1 + \frac{\mathbf{e}_\alpha \cdot \mathbf{u}}{c_s^2} + \frac{(\mathbf{e}_\alpha \cdot \mathbf{u})^2}{2c_s^4} - \frac{(\mathbf{u} \cdot \mathbf{u})}{c_s^2} \right] \quad (18)$$

Furthermore, in Eq. (12), $F_\alpha(\mathbf{x}, t)$ is the forcing term, which may be defined by [43]:

$$F_\alpha(\mathbf{x}, t) = \delta t \omega_\alpha \frac{\mathbf{e}_\alpha \cdot \mathbf{F_t}}{\rho c_s^2} \quad (19)$$

where $\mathbf{F_t} = \mathbf{F_\mu} + \mathbf{F_p} + \mathbf{F_s} + \mathbf{F_b} + \mathbf{F_m}$ represents the summation of applied forces, in which $\mathbf{F_\mu} = \nu[\nabla \mathbf{u} + (\nabla \mathbf{u})^T] \cdot \nabla \rho$ is the viscous force and $\mathbf{F_p} = \frac{-p}{\rho} \nabla \rho$ is the pressure force [47]. The implemented forcing term has been suggested and verified by Zu and He [43]. In addition, Fakhari et al. [47] stated that in multiphase lattice-Boltzmann models, utilizing the higher-order forms of the forcing term does not alter the results. It should be expressed that the magnetic force $\mathbf{F_m}$ is determined by FDM (see section 2.1). Moreover, in the D2Q9 model, the discrete particle velocities $\mathbf{e}_\alpha$ are [50, 51]:

$$\mathbf{e}_\alpha = \begin{cases} (0,0), & \alpha = 0 \\ (cos[(i-1)\pi/2], sin[(i-1)\pi/2])c, & \alpha = 1-4 \\ (cos[(2i-9)\pi/4], sin[(2i-9)\pi/4])\sqrt{2}\,c, & \alpha = 5-8 \end{cases} \quad (20)$$

where $c = \delta x/\delta t$ is the lattice speed, while $\delta x$ is the steaming length.

Ultimately, once Eq. (12) is solved, the pressure and velocity fields can be accomplished as follows, respectively:

$$p = \rho c_s^2 \sum_a f_\alpha \quad (21)$$

$$\mathbf{u} = \sum_a f_\alpha \mathbf{e}_\alpha + \frac{\mathbf{F_t}}{2\rho} \delta t \quad (22)$$

It is worth mentioning that the distribution function $f_\alpha$ is dimensionless. Furthermore, for the sake of simplicity, we take $\delta x = 1$ and $\delta t = 1$ in all simulations.

## 2.3 A conservative phase-field model for the interface

In the phase-field model, in lieu of abrupt changes across the interface of various fluids involved, the thermophysical properties of the fluid are gradually altered in the transition region [52]. In fact, two phase-field theories have been introduced, namely Cahn–Hilliard (C–H) [53] and Allen–Cahn (A–C) [40]. Although the A-C equation has several advantages over the C-H equation [41, 54], the selection between the A-C and C-H models should be accomplished based on the specific phenomena being considered [55]. In this research, in light of the physics involved in the problem [39], the conservative AC model suggested by Sun and Beckermann [56] and amended by Chiu and Lin [57] is exploited. Moreover, the implementation of the conservative A-C equation is more convenient than the C-H equation since a second-order algorithm is required so that the diffusion term can be discretized. The conservative A-C equation can be delineated as follows [52]:

$$\frac{\partial \phi_i}{\partial t} + \nabla \cdot (\mathbf{u}\phi_i) = M_i \nabla \cdot \left[ \nabla \phi_i - \mathbf{n}_i \frac{4}{\xi} \phi_i(1-\phi_i) + \frac{1}{3}\sum_{i=1}^{3} \mathbf{n}_i \frac{4}{\xi} \phi_i(1-\phi_i) \right] \quad (23)$$

where $M_i$ indicates the mobility of fluid $i$. Also, the interface width is symbolized by $\xi$. Presumed to be the same for all phases, $\xi = 5$ lattice units is adopted for all simulations [58]. Furthermore, $\mathbf{n}_i = \frac{\nabla \phi_i}{|\nabla \phi_i|}$ denotes the unit vector normal to the interface of fluid $i$. The ternary system governed by Eq. (23) should comply with two paramount constraints for which the reader is referred to the article of Haghani Hassan Abadi et al. [52].

It is worth mentioning that the phase-field variables satisfy the following relation:

$$\phi_1 + \phi_2 + \phi_3 = 1 \quad (24)$$

which conveys that due to the conservation of mass, merely two of the phase-field variables should be computed, while the third one is attained through Eq. (24).

Also, the chemical potential ($\mu_{\phi_i}$) of each phase may be procured by [59]:

$$\mu_{\phi_i} = \frac{4\gamma_T}{\xi} \sum_{j \neq i} \left[ \frac{1}{\gamma_j} \left( \frac{\partial \mathcal{H}}{\partial \phi_i} - \frac{\partial \mathcal{H}}{\partial \phi_j} \right) \right] - \frac{3}{4} \xi \gamma_t \nabla^2 \phi_i \quad (25)$$

where $\gamma_T$ can be achieved by the following relation:

$$\frac{3}{\gamma_T} = \frac{1}{\gamma_1} + \frac{1}{\gamma_2} + \frac{1}{\gamma_3} \tag{26}$$

where $\gamma_i$ is an auxiliary coefficient, whose definition is provided below:

$$\gamma_i = \sigma_{ij} + \sigma_{ik} - \sigma_{jk}, \quad i, j, k = 1, 2, \text{ and } 3 \tag{27}$$

where $\sigma_{ij}$ represents the surface tension between phases $i$ and $j$.

Moreover, in Eq. (25), $\mathcal{H}$ denotes the bulk free energy, which is described as follows [60]:

$$\mathcal{H}(\phi_1, \phi_2, \phi_3) = \sigma_{12}\phi_1^2\phi_2^2 + \sigma_{13}\phi_1^2\phi_3^2 + \sigma_{23}\phi_2^2\phi_3^2 + \phi_1\phi_2\phi_3(\gamma_1\phi_1 + \gamma_2\phi_2 + \gamma_3\phi_3) + \eta\phi_1^2\phi_2^2\phi_3^2 \tag{28}$$

where the constant parameter $\eta$ is zero for partial spreading, while it is nonzero for total spreading. Through the spreading parameter $S_i = -\gamma_i$, one may differentiate between partial and total spreading. Indeed, when $S_i > 0$ for at least one phase, the total spreading condition is acquired. On the contrary, as $S_i < 0$ for all phases, the partial spreading case is achieved [58].

## 2.4 LBM for the interface tracking

An additional distribution function is employed so that Eq. (23) can be solved by the LBM [52]:

$$g_\alpha^i(\mathbf{x} + \mathbf{e}_\alpha \delta t, t + \delta t) = g_\alpha^i(\mathbf{x}, t) - \frac{g_\alpha^i(\mathbf{x}, t) - g_\alpha^{i,eq}(\mathbf{x}, t)}{\frac{1}{2} + \tau_\phi} + F_\alpha^{\phi,i}(\mathbf{x}, t) \tag{29}$$

where $g_\alpha^i$ and $g_\alpha^{i,eq}$ are the phase-field distribution function and phase-field equilibrium distribution function of fluid $i$, respectively. Also, $\tau_\phi$ is phase-field relaxation time, which can be determined as follows [47]:

$$\tau_\phi = \frac{M}{c_s^s \delta t} \tag{30}$$

where the mobility $M$ is presumed to be constant for all fluids ($M = 0.3$). Having been specified by numerous trials and errors, this mobility value ensures the stability and accuracy of simulations within the range over which parameters vary in this research. In addition, $F_\alpha^{\phi,i}(\mathbf{x}, t)$ is the representative of the source term whose definition is provided in Ref. [39]. Moreover, the phase-field equilibrium distribution function of the fluid $i$ ($g_\alpha^{i,eq}$) may be obtained as follows [52]:

$$g_\alpha^{i,eq} = \phi \Gamma_\alpha - \frac{1}{2} F_\alpha^{\phi,i} \tag{31}$$

where $\Gamma_\alpha$ was introduced earlier. When Eq. (29) is solved via the conventional collision and streaming procedure, the phase-field variables can be attained by:

$$\phi_i = \sum_\alpha g_\alpha^i \quad i = 1, 2 \tag{32}$$

Also, based on Eq. (24), $\phi_3 = 1 - \phi_1 - \phi_2$ [52]. Afterward, density can be calculated by linear interpolation:

$$\rho = \sum_{i=1}^{3} \rho_i \phi_i \tag{33}$$

Subsequently, the gradient and Laplacian of the phase-field variables may be attained through the following equations, respectively:

$$\nabla \phi_i = \frac{c}{c_s^2 \delta x} \sum_\alpha \mathbf{e}_\alpha \omega_\alpha \phi_i(\mathbf{x} + \mathbf{e}_\alpha \delta t, t) \tag{34}$$

$$\nabla^2 \phi_i = \frac{2c^2}{c_s^2 (\delta x)^2} \sum_\alpha \omega_\alpha [\phi_i(\mathbf{x} + \mathbf{e}_\alpha \delta t, t) - \phi_i(\mathbf{x}, t)] \tag{35}$$

The unit vector normal to the interface ($\mathbf{n}_i = \frac{\nabla \phi_i}{|\nabla \phi_i|}$) can be determined by the gradient of the phase-field variable [Eq. (2-34)]. Ultimately, in the MRT-LBM formulation, the surface tension force takes the following form [59]:

$$\mathbf{F_s} = \sum_{i=1}^{3} \mu_{\phi_i} \nabla \phi_i \tag{36}$$

The algorithm through which the hybrid LBM-FDM approach is executed for characterizing the ternary flow system and the magnetic field is demonstrated in Fig. 1.

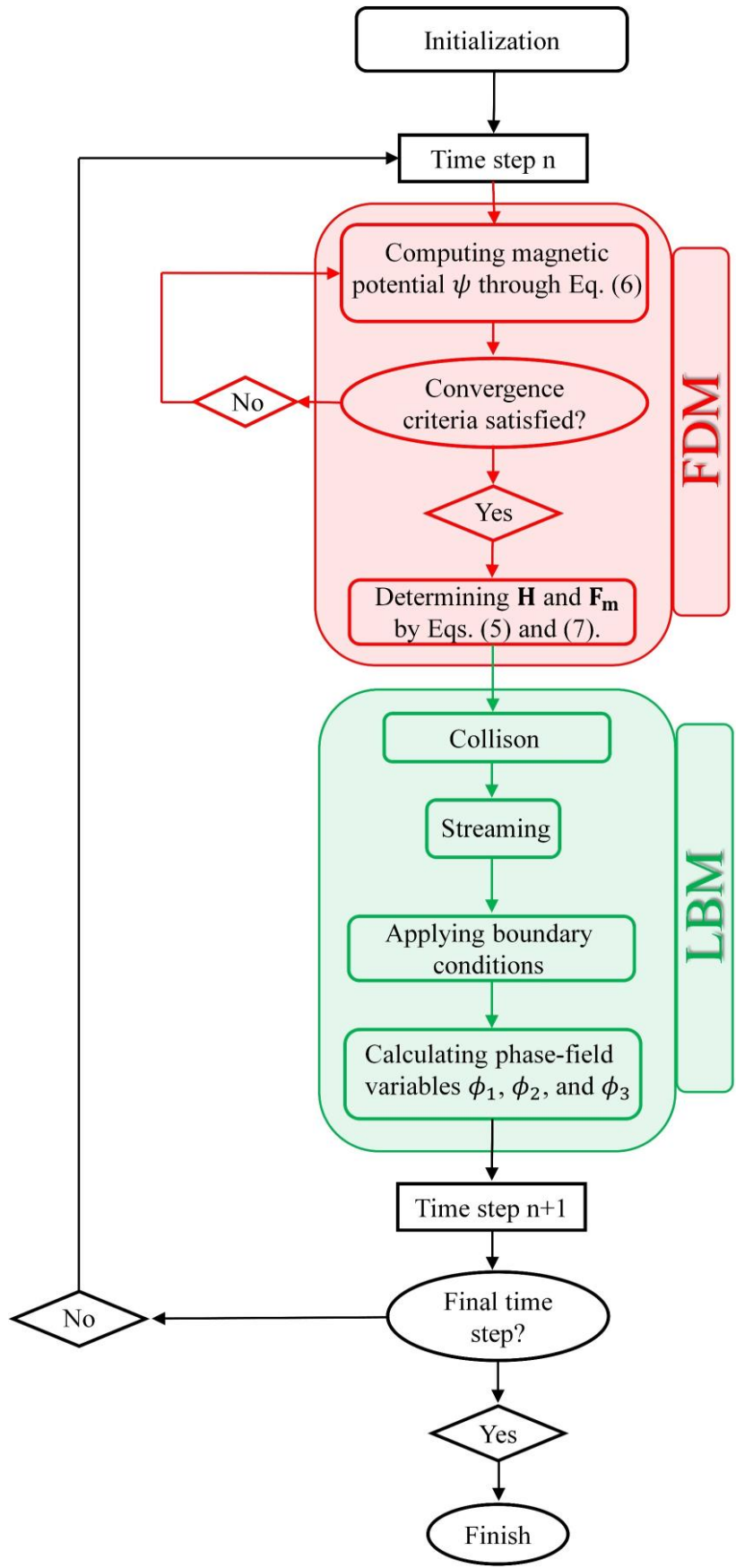

Fig. 1 The utilized numerical algorithm to fulfill hybrid LBM-FDM simulations.

## 2.5 Problem definition

Fig. 2 reveals the configuration of a single-core FCD in a horizontal microchannel under the simultaneous impacts of the pressure-driven flow and a UEMF. The UEMF, which is imposed at the angle $\alpha$ relative to the horizontal axis, has a constant strength of $H_0$. The inner and outer droplets bear the radius of $r_1$ and $r_2$, respectively. The investigation is conducted in two scenarios, in the first of which the inner droplet is supposed to be ferrofluid, whereas, in the second scenario, the outer droplet is ferrofluid. In each scenario, the influence of the magnetic field on the deformation and breakup of the compound droplet is punctiliously surveyed. At the inception of the computation, the compound droplet, whose core and shell are spherical and concentric, is located at the distance of $4r_1$ from the channel inlet. Moreover, the compound droplet center is coincident with the channel center line. The microchannel length and height are $L$ and $H$, respectively. All fluids are presumed to be Newtonian, incompressible, and immiscible. Throughout the text, the subscripts 1, 2, and 3 refer to the inner droplet (core), outer droplet (shell), and ambient fluid, respectively.

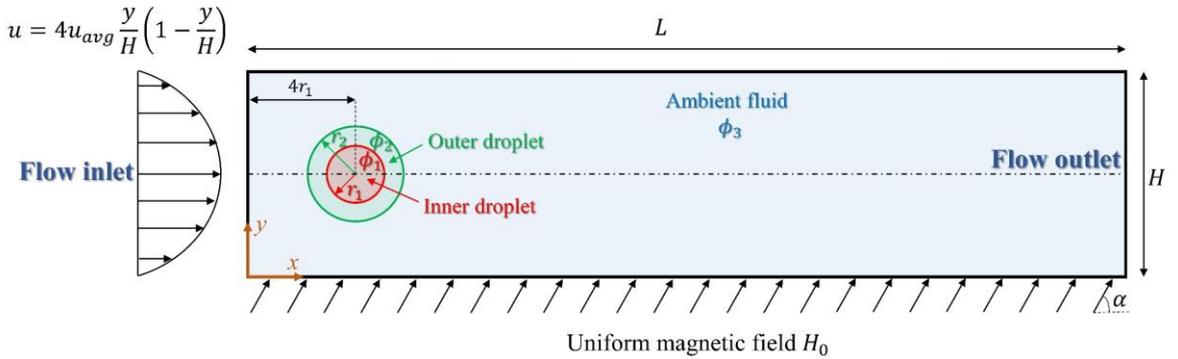

Fig. 2 Schematic of the initial configuration of an FCD moving in a microchannel.

The bounce-back boundary condition is applied to the top and bottom walls so that the no-slip boundary may be achieved [45]. In addition, the inlet velocity condition with a fully developed laminar flow is brought to bear on the left side according to Zu and He [43] model as follows:

$$\begin{cases} u = 4u_{avg}\dfrac{y}{H}\left(1 - \dfrac{y}{H}\right) \\ v = 0 \end{cases} \tag{37}$$

where $u_{avg}$ is the average inlet velocity. Furthermore, having a promising accuracy in binary [61] and ternary [62] fluids, the convective boundary condition (CBC) is imposed on the right side as the outlet boundary condition.

Herein, the outer droplet diameter is considered as the characteristic length; thus, the dimensionless time can be defined as

$$t^* = \frac{tu_{avg}}{2r_2} \tag{38}$$

in which $t$ is the physical time. The dimensionless parameters via which the dynamics of the problem may be delineated are provided below:

$$r_r = \frac{r_2}{r_1}, \rho_r = \frac{\rho_1}{\rho_2} = \frac{\rho_3}{\rho_2}, v_r = \frac{v_1}{v_2} = \frac{v_3}{v_2} \tag{39}$$

where $r_r$, $\rho_r$, and $v_r$ are the ratios of radius, density, and kinematic viscosity, respectively. Furthermore, specifying the ratio of viscous force to the interfacial tension force, the Ca number can be determined by

$$Ca = \frac{\rho_3 v_3 u_{avg}}{\sigma_{23}} \tag{40}$$

where $\sigma_{23}$ symbolizes the surface tension between the shell and ambient fluid. Moreover, representing the ratio of the magnetic force to the interfacial tension force, the Bo$_m$ number may be described as follows depending on which droplets (core or shell) are treated as ferrofluid:

$$Bo_m = \begin{cases} \frac{\mu_0 \chi_1 H_0^2 r_1}{2\sigma_{12}}, \text{ inner droplet is ferrofluid} \\ \frac{\mu_0 \chi_2 H_0^2 (r_2^2 - r_1^2)}{2(\sigma_{12} r_1 + \sigma_{23} r_2)}, \text{ outer droplet is ferrofluid} \end{cases} \tag{41}$$

where $\chi_1$ and $\chi_2$ are set to unity whenever the inner and outer droplets are ferrofluid, respectively; otherwise, they are fixed at zero. $\chi_3 = 0$ in all simultasions as well. Also, $\sigma_{12}$ indicates the surface tension between the inner and outer droplets. Table 1 summarizes the ranges at which the mentioned parameters vary in this research.

Table 1 Variations of parameters exploited in this research.

| Parameter | Description | Range |
|---|---|---|
| $\alpha$ | Magnetic field angle | 0° and 90° |
| $r_r$ | Radius ratio | 1.4 – 2.3 |
| $\rho_r$ | Density ratio | 1 - 10 |
| $\nu_r$ | Viscosity ratio | 1 - 10 |
| Ca | Capillary number | 0.05 – 0.30 |
| Bo$_m$ | Magnetic Bond number | 0 – 8.0 |
| $\sigma_{13}/\sigma_{12}$ | Surface tension ratio | 0.3 – 1.7 |

To evaluate the FCD deformation prior to the rupture quantitatively, the following relation is utilized [63]:

$$T_o = \frac{\ell_{y_o} - \ell_{x_o}}{\ell_{y_o} + \ell_{x_o}} \qquad (42)$$

where $T_o$ is the deformation index of the outer droplet, $\ell_{y_o}$ is the maximum distance between the upper and lower points of the outer droplet interface, whereas $\ell_{x_o}$ indicates the maximum distance between the left and right front of the outer droplet interface, located in channel centerline. $T_i$, $\ell_{y_i}$, and $\ell_{x_i}$ are defined similarly. Given Eq. (39), when $T_o < 0$, the shell stretches in the $x$ direction, resulting in an oblate-like shape. On the contrary, for $T_o > 0$, the shell elongates in the $y$ direction, giving rise to a prolate-like configuration, which is also the case for the inner droplet. Herein, the breakup is defined as the condition under which the inner droplet touches the interface of the shell. Fig. 3 displays the nondimensional numbers associated with the deformation and breakup phenomena, in which $t_b^*$ is the nondimensional time at which the breakup occurs. $y_b^*$ is a dimensionless parameter quantifying the distance between the channel centerline and the breakup point. Additionally, nondimensionalized by the channel height, $L_b^*$ represents the breakup location with respect to the $y$-axis ($x = 0$). In this research, the terms rupture and breakup are employed interchangebly.

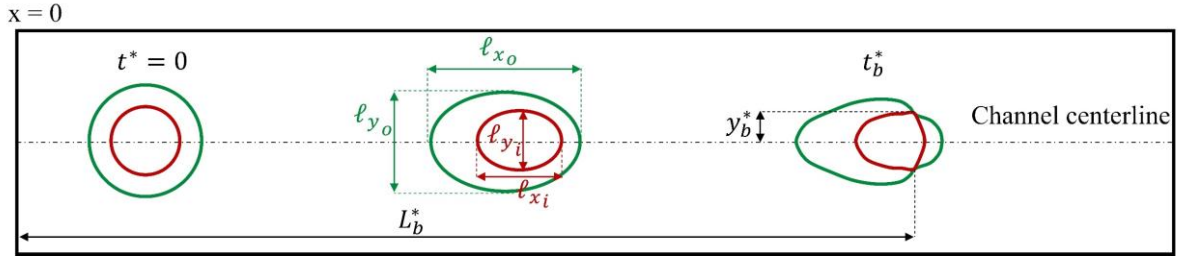

Fig. 3 Definition of the nondimensional numbers associated with the FCD deformation and breakup.

The physics involved in this problem includes the interaction between three phases, the deformation of a simple ferrofluid droplet under UEMF, and droplet deformation in fluid flow. To verify the accuracy of the developed solver, all of these physics were validated separately using existing literature in our previous work [39]. Hence, the validation of the numerical approach is skipped in this paper to avoid superfluous explanations.

## 2.6 Grid convergence

As alluded to previously, the domain size is $L \times H$, in which the channel length ($L$) is six times as high as the channel height ($H$). Additionally, compared to the shell diameter, $L$ is selected so lengthy that the channel outlet may not affect the compound droplet migration and rupture processes. To ensure grid independence, three grid resolutions of 100 × 600, 150 × 900, and 200 × 1200 are explored. Fig. 4 compares the grid refinement results when the shell is ferrofluid, Ca = 0.15, $\alpha = 0°$, and $r_{\bar{r}} = 1.75$, while the other ratios are fixed at 1. The convergence results encompass the variations of the outer droplet deformation ($T_o$) with respect to the magnetic Bond number [Fig. 4 (a)] when $t^* = 1.84$, the compound droplet shape at $Bo_m = 2$ [Fig. 4 (b)] when $t^* = 1.84$, and the migration process of the FCD at distinct moments for different grids [Fig. 4 (c)]. The difference between the two finer grids (150 × 900 and 200 × 1200) is, on average, 5.2%. Conversely, the results yielded from the coarse grid (100 × 600) significantly deviate from the finest one with an average error of 17.8%. Therefore, from now on, the grid size of 150 × 900 is utilized for the simulations, which reduces the computational time compared to the finest grid, although not influencing the accuracy of the results. The grid convergence is also explored for distinct values of the Ca and $Bo_m$ numbers, corresponding to the distinct regimes observed during the transport of an FCD in the microchannel under the UEMF (see section 3.7), whose results are given in Fig. S2 of the Supplementary material.

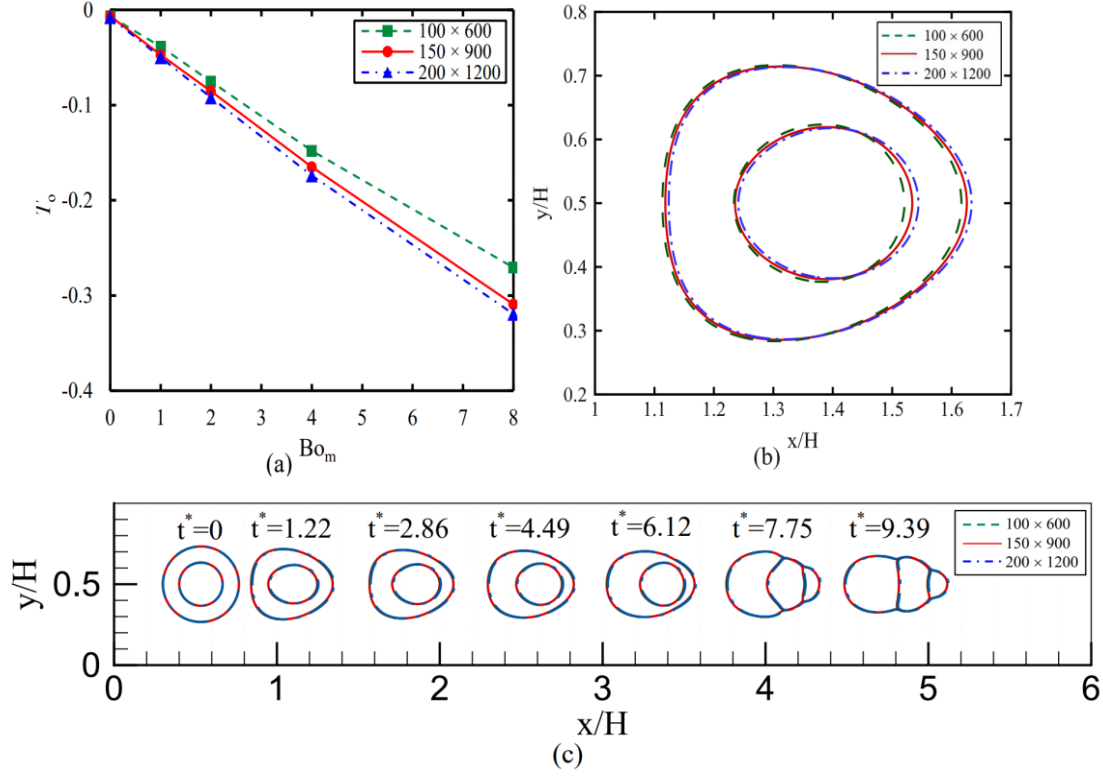

Fig. 4 Comparison of the results procured by distinct grid resolutions to ascertain the foremost grid size for the rest of the simulations; (a) the variations of the outer droplet deformation ($T_o$) with the magnetic Bond number, (b) compound droplet shape when transporting in the channel at $Bo_m = 2$, and (c) the migration process of an FCD in the microchannel at various moments. The results are presented at $t^* = 1.84$ when Ca = 0.15, $\alpha = 0°$, $r_r = 1.75$, $\rho_r = 1$, $\nu_r = 1$, and the shell is treated as ferrofluid.

## 3  Results and discussion

In this section, the impacts of the influential parameters on the deformation and breakup of an FCD moving in a microchannel under the concurrent existence of the background pressure-driven flow and the magnetic field are assessed systematically. Initially, section 3.1 is dedicated to scrutinizing the influence of the magnetic Bond number on the FCD deformation dynamics depending on whether the core or shell is ferrofluid for two different magnetic field angles of 0° and 90°. Afterward, the effects of Capillary number, density ratio, viscosity ratio, radius ratio, and surface tension coefficients are investigated in sections 3.2 to 3.5, respectively. Ultimately, section 0 is allocated to analyzing the compound droplet morphology and distinguished regimes occurring at the rupture moment. Unless otherwise stated, simulations are fulfilled for Ca = 0.15, $\rho_r = 1$, $\nu_r = 1$, $r_r = 1.75$, and $\sigma_{12} = \sigma_{23} = \sigma_{13}$. In addition, the radius of inner and outer droplets are fixed at 20 and 35 lattice units, respectively. Furthermore, as a case study, the hydrodynamics of two conventional configurations of compound droplets, namely oil-in-water-in oil (O/W/O) and water-in-oil-in-water (W/O/W), are also investigated, whose results are provided in the supplementary material. The paramount purpose of the following results is to alter the governing parameters such that the rupture is postponed, which is indispensable in the targeted drug delivery application.

### 3.1 Impact of the magnetic Bond number

#### 3.1.1 The core is ferrofluid

Fig. 5 (a) and (b) reveal the temporal evolution of the inner and outer droplets' deformation, respectively, for various $Bo_m$ numbers in advance of the rupture, in which the inner droplet is assumed to be ferrofluid ($\chi_1 = 1$) and $\alpha = 0°$. Also, Fig. 5 (c) indicates the variations of $t_b^*$, $L_b^*$, and $y_b^*$ with respect to the $Bo_m$ number. Fig. 6 compares the configuration of the FCDs when transporting in the microchannel, where the solid red and solid green lines are indicative of the interface of inner and outer droplets, respectively. In fact, in the represented contours, the interface is indicated at $\phi_i = 0.5$. It should be expressed that the compound droplet shape after the breakup is also encompassed in the represented contours so that readers may gain a more exhaustive insight into its morphology.

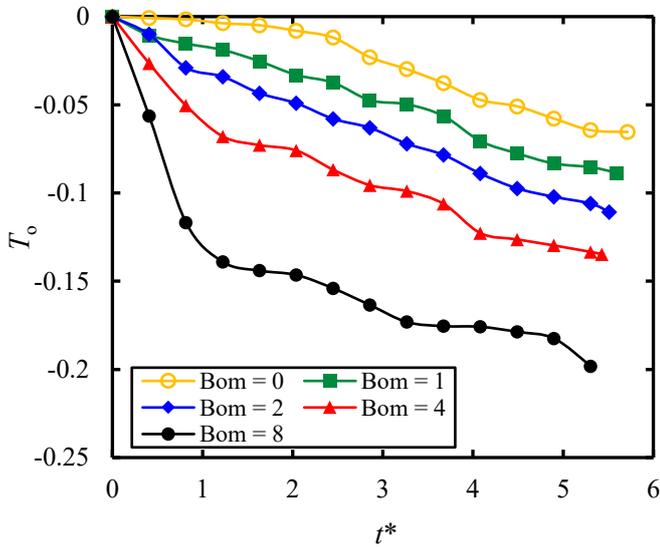

(a)

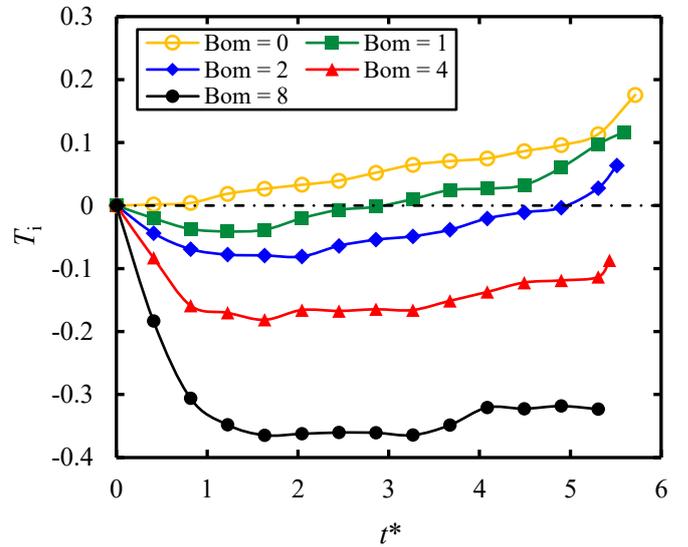

(b)

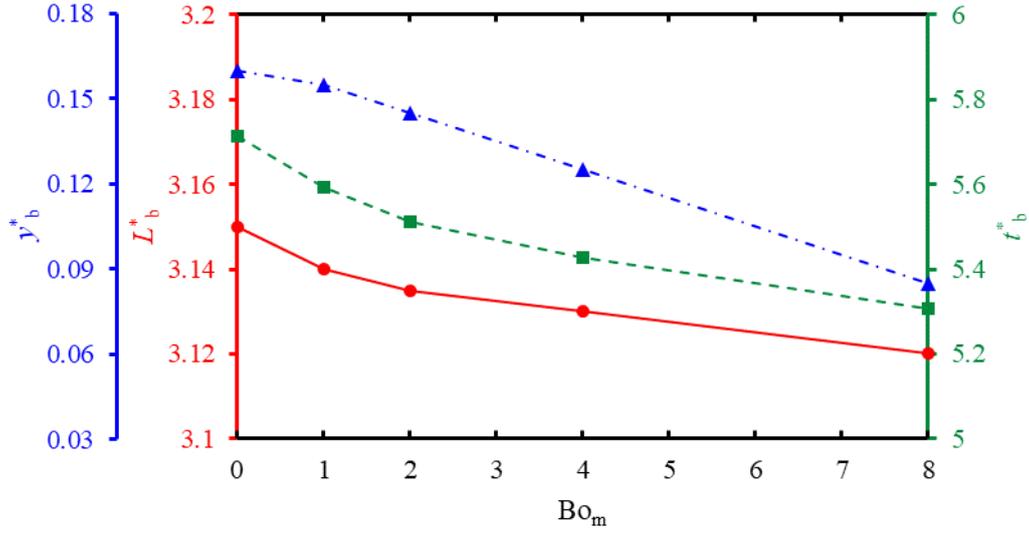

(c)

Fig. 5 Influence of the $Bo_m$ number on the FCD deformation and breakup over the course of traversing in a microchannel; (a) variation of the shell deformation with respect to the dimensionless time, (b) variation of the inner droplet deformation with dimensionless time, and (c) $L_b^*$, $t_b^*$, and $y_b^*$ as a function of the magnetic Bond number. The inner droplet is supposed to be ferrofluid and $\alpha = 0°$.

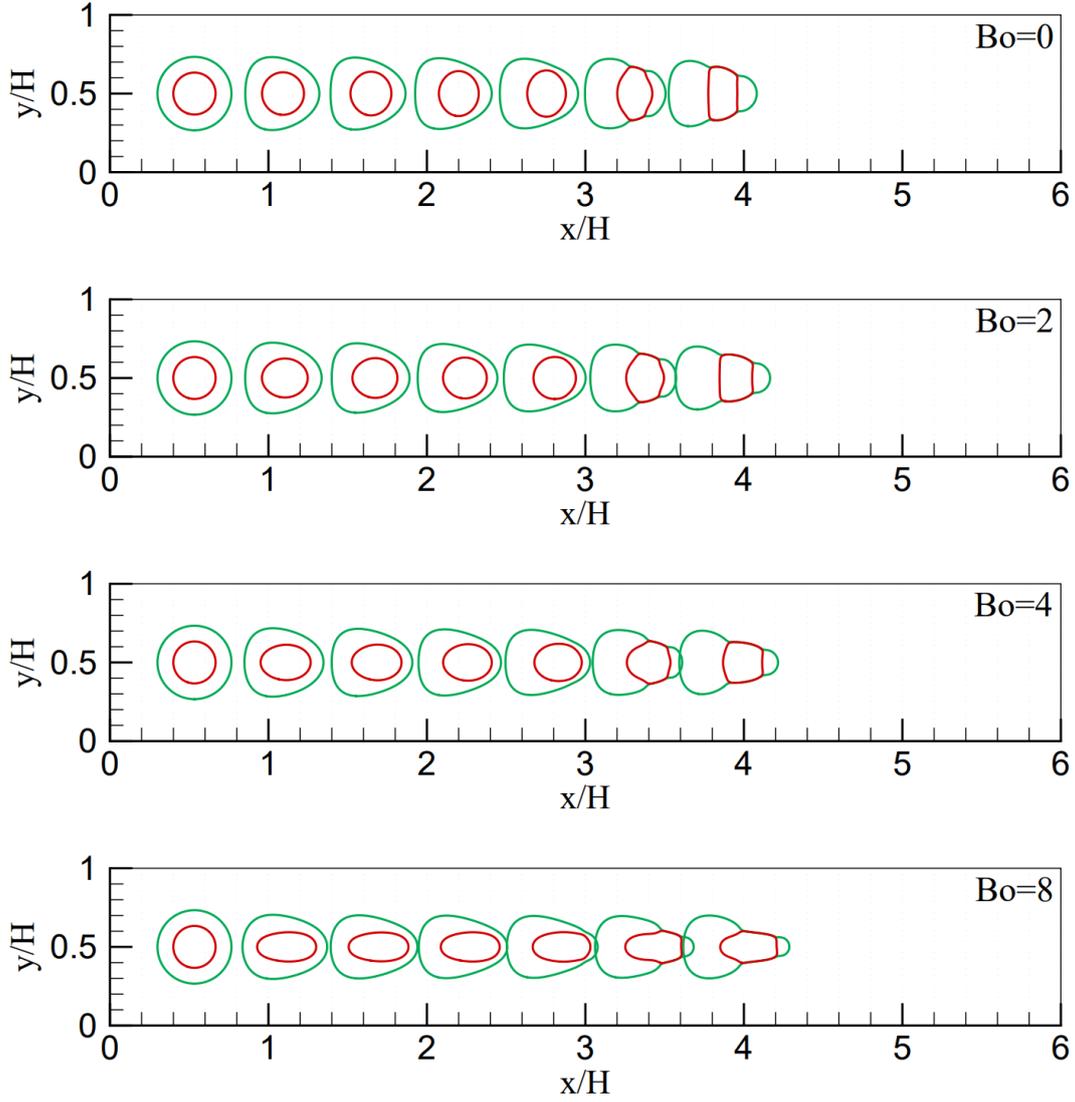

Fig. 6 The migration of an FCD inside a microchannel from the initial point to the outlet over time for various magnetic Bond numbers at $\alpha = 0°$ when the inner droplet is ferrofluid. From left to right: $t^* =$ 0, 1.22, 2.45, 3.67, 4.9, 6.12 and 7.35.

Regarding Fig. 5 (a) and Fig. 5 (b), at Bo = 0 in which the magnetic field is not imposed, the compound droplet migrates under the sole effect of the background pressure-driven flow, inducing a drag force, which elongates the outer and inner droplets along the *x* and *y* axes, corresponding to the negative and positive values of the deformation parameter, respectively, as shown in Fig. 6. To put it another way, the outer droplet stretches along the flow direction, engendering an oblate-like shape, while the inner droplet deforms perpendicular to the flow, provoking a prolate-like shape, which is consistent with the results provided by Santra et al. [64]. Since experiencing a lower drag force owing to its smaller size, the inner droplet moves more swiftly than the shell, implying that the compound droplet deforms eccentrically during traversing in the microchannel, resulting in the breakup phenomenon at $t_b^* = 5.71$.

Additionally, when $Bo_m = 0$, the deformation of the inner droplet is, on average, 57.0% greater than that of the outer droplet, arguably emanating from the difference in the drag force applied to them.

Based on Fig. 5 and Fig. 6, once applied, the external magnetic field drastically alters the deformation of the FCD, conveying that by deploying the magnetic field, one may cause or hinder the changes in the FCD morphology during its movement in a pressure-driven flow. It is worth mentioning that when the inner droplet is assumed to be ferrofluid, the magnetic force is imposed solely on the interface of the inner droplet, whose diagram is illustrated and discussed later in this section. In other words, the magnetic force is not applied to the outer droplet interface since it is merely imposed on the interfaces where $\nabla \chi \neq 0$ [see Eq. (7)]. In the case of the outer droplet, the drag force elongates the shell in the *x* direction and deforms its rear side to a flat shape [see Fig. 6]. Interestingly, although the magnetic force is not brought to bear on the shell interface, $T_o$ rises with the $Bo_m$ number increment, as evident in Fig. 5 (a). In fact, the magnetic force stretches the inner droplet horizontally, compressing the fluid in the frontal side of the shell, which exerts an outward force on the shell interface, leading to its further horizontal elongation in the frontal side. Given the inner droplet, the drag and magnetic forces compete to elongate the core along and perpendicular to the flow direction, respectively, justifying why $T_i$ is not raised monotonously with time and minimum points are observed for the $T_i$ diagrams in Fig. 5 (b). Indeed, at the inception of the migration, the magnetic force brings about an oblate-like shape; however, the drag force gradually reduces the deformation magnitude, axiomatic in Fig. 5 (b). When Bo < 4, the viscous force eventually surpasses the magnetic force, deforming the inner droplet into a prolate-like shape ($T_i > 0$) at the rupture moment. Conversely, at Bo ≥ 4, the magnetic force prevails over the viscous force; hence, the inner droplet preserves its oblate-like configuration ($T_i < 0$). By augmenting the magnetic Bond number, inasmuch as chosen to be ferrofluid, the inner droplet undergoes much more deformation than the outer droplet, as obvious in Fig. 5.

Fig. 5 (c) suggests that increasing the magnetic Bond number slightly reduces $t_b^*$ and $L_b^*$. Actually, by strengthening the magnetic field, since ferrofluid, the inner droplet stretches drastically while the outer droplet is less elongated along the *x*-axis, gradually diminishing the distance between the core and shell, which gives rise to the rupture phenomenon. Accordingly, the higher the $Bo_m$ number, the sooner the breakup and the shorter the breakup length ($L_b^*$), from which one may infer that selecting the inner droplet as the ferrofluid and applying the magnetic field at $\alpha = 0°$ would be unfavorable for postponing the rupture process. Furthermore, concerning Fig. 5 (c), $y_b^*$ declines with the magnetic Bond number since by intensifying the magnetic field, the frontal side of the outer droplet becomes narrow and prolonged, lowering the height at which the inner droplet contacts the shell at the rupture instant.

The time evolution of the outer and inner droplet deformations prior to the rupture is depicted in

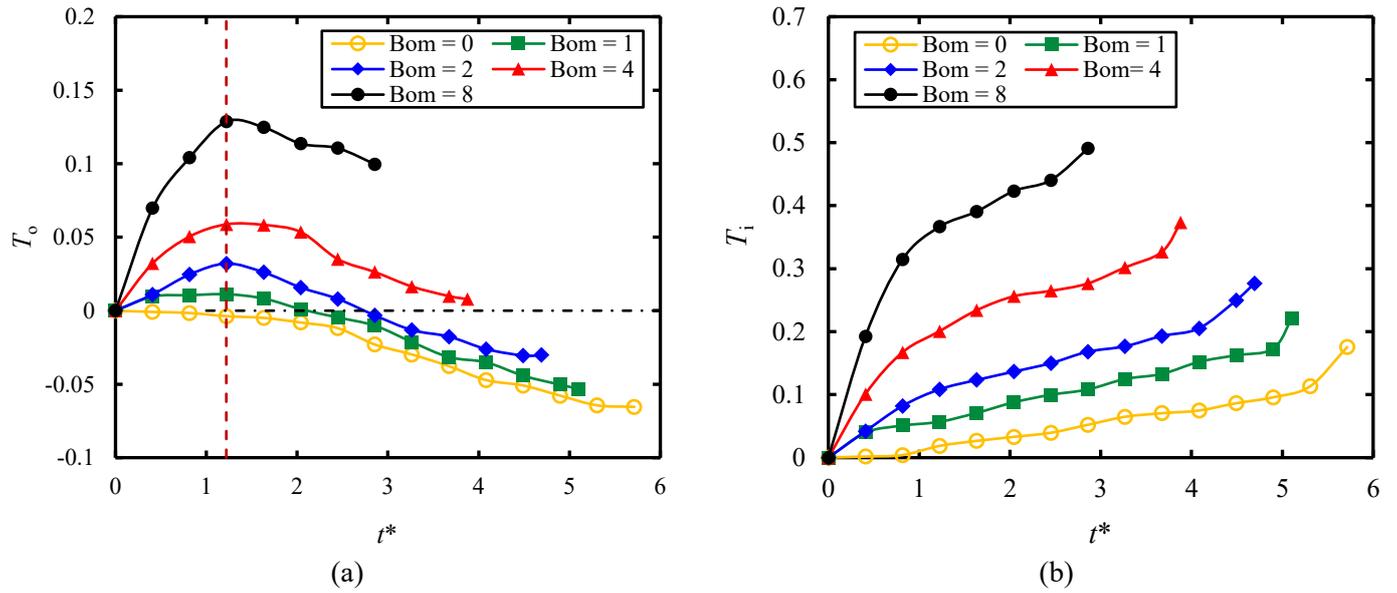

(a)

(b)

(c)

Fig. 7 for disparate $Bo_m$ numbers, in which the inner droplet is assumed to be ferrofluid and $\alpha = 90°$. The transport process of the FCD is illustrated in Fig. 8. For the outer droplet [

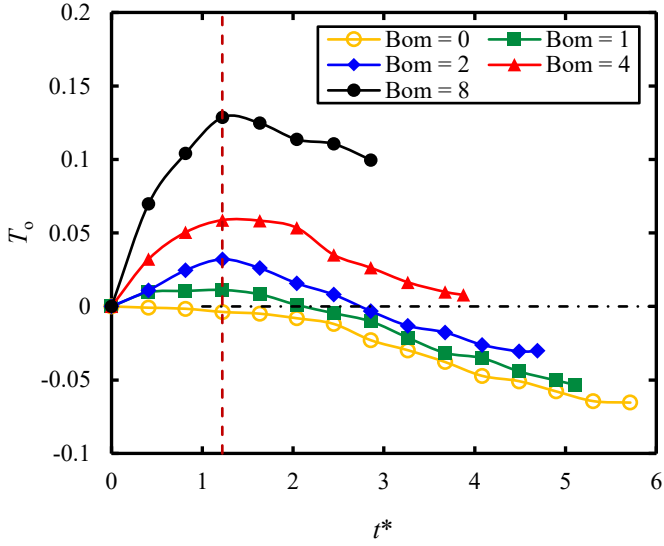 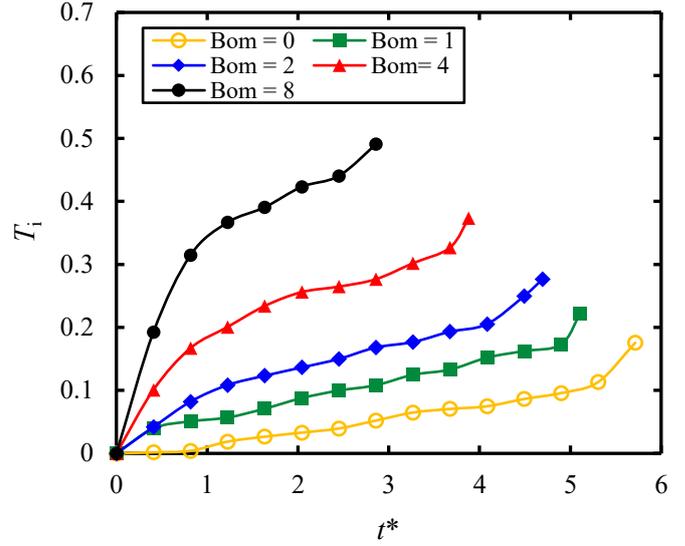

(a)

(b)

(c)

Fig. 7 (a)], in contrast to the previous case discussed above, the drag force and the force generated by the core vertical deformation endeavor to deform the shell along $x$ and $y$ directions, respectively. This incidence elucidates the reason behind the peak points observed in $T_o$ graphs. Initially, due to the core elongation, the fluid in the upper and lower sides of the shell compresses, generating a vertical outward force, which in turn deforms the outer droplet into a prolate-like shape. However, by virtue of the drag force, outer droplet deformation lessens. When Bo ≥ 4, the force caused by the core deformation dominates the drag force, and the outer droplet maintains the prolate-like configuration at the breakup instant. On the contrary, for Bo < 4, the viscous force ultimately deforms the outer droplet into an oblate-like shape, corresponding to $T_o < 0$ (see Fig. 8). Intriguingly, the time at which $T_o$ accomplishes the peak point and commences to decline is identical for various $Bo_m$ numbers, which is shown by $t_c^*$ in

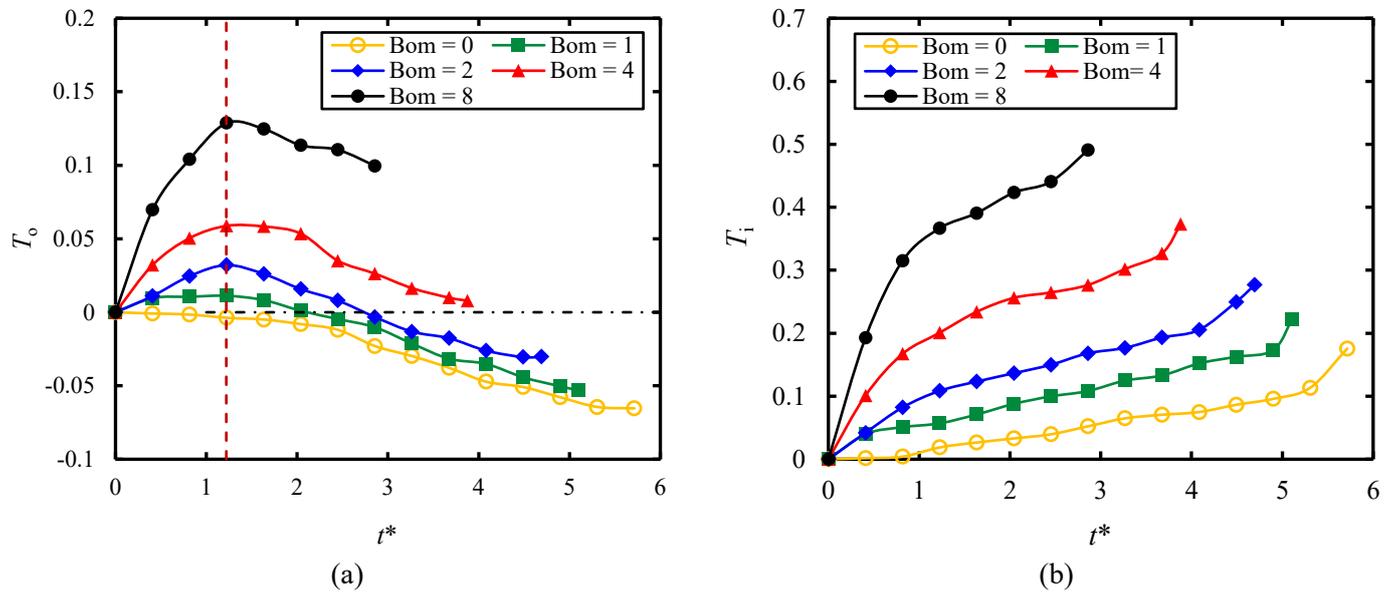

(a)

(b)

(c)

Fig. 7 (a). Generally speaking, the outer droplet deformation is meager ($T_o < 0.15$) for all values of the $Bo_m$ number, proposing that when not ferrofluid, the outer droplet undergoes a finite deformation. In the case of the inner droplet [

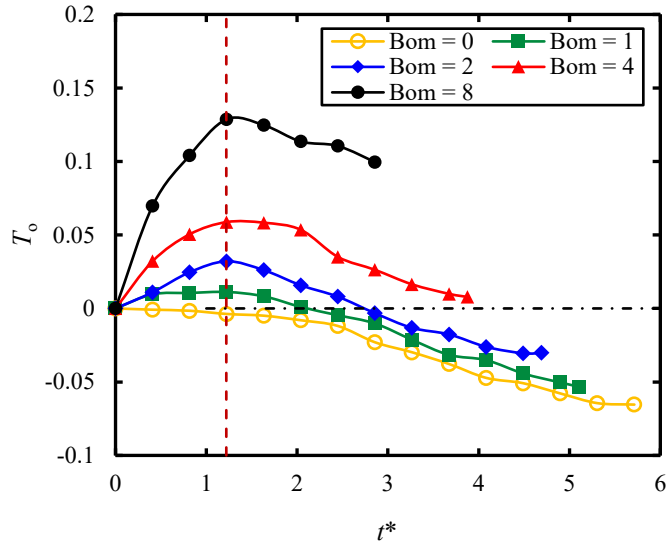
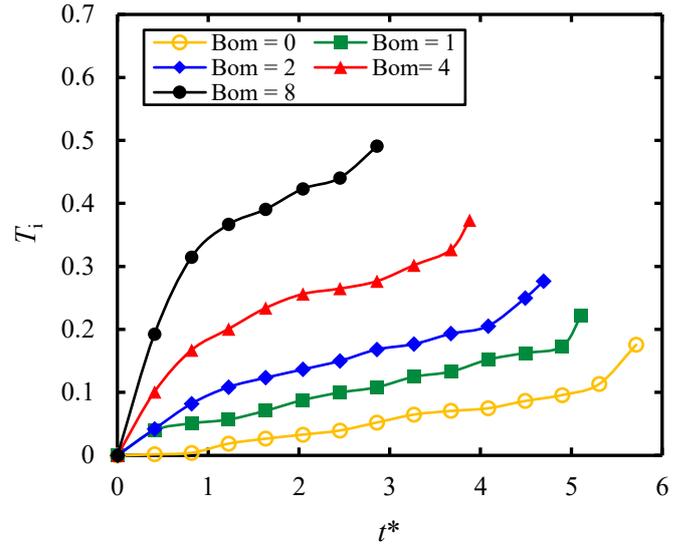

(c)

Fig. 7 (b)], magnetic and drag forces are exerted perpendicular to the fluid flow, both of which stretch the inner droplet along the $y$ direction ($T_i > 0$), which is evident in Fig. 8 as well. Moreover, $T_i$ exceedingly boosts with the magnetic Bond number. Contemplating

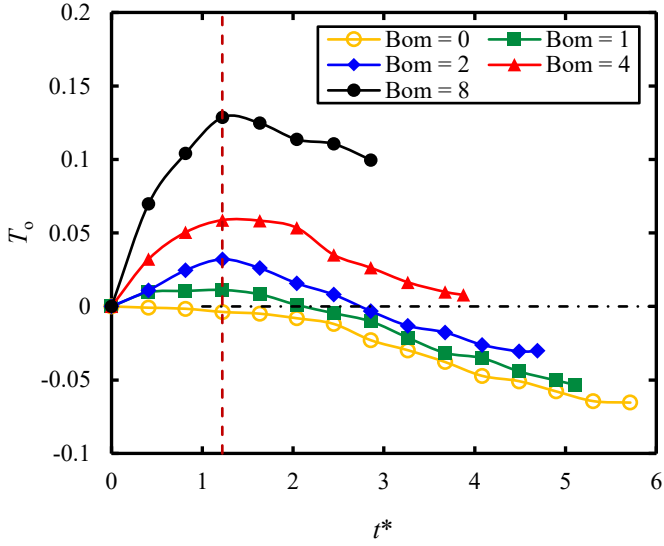
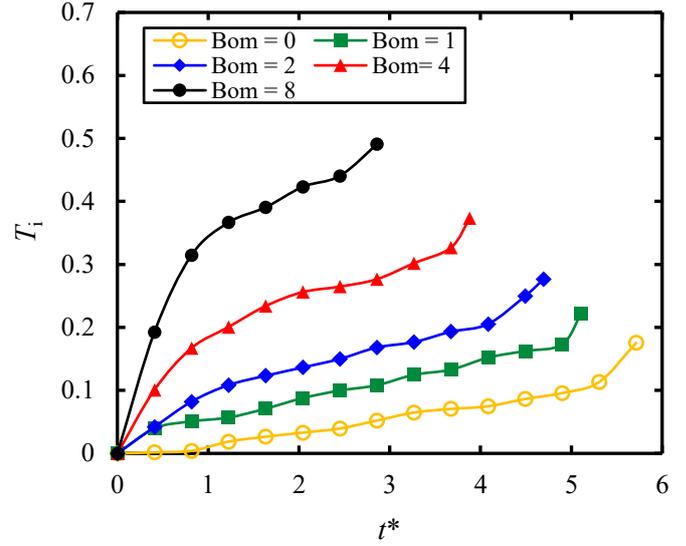

(a)

(b)

(c)

Fig. 7 (c), at $\alpha = 90°$, increasing the $Bo_m$ number precipitates the rupture process drastically. Quantitatively speaking, raising the $Bo_m$ number from 1 to 8 decreases the breakup time by 44.0%. In fact, as elongated vertically, the inner droplet does not possess sufficient space to move inside the shell, and it promptly touches the interface of the shell, instigating the breakup incidence and curtailing $t_b^*$ and $L_b^*$. This rationale also accounts for $y_b^*$ increment as the $Bo_m$ number enlarges.

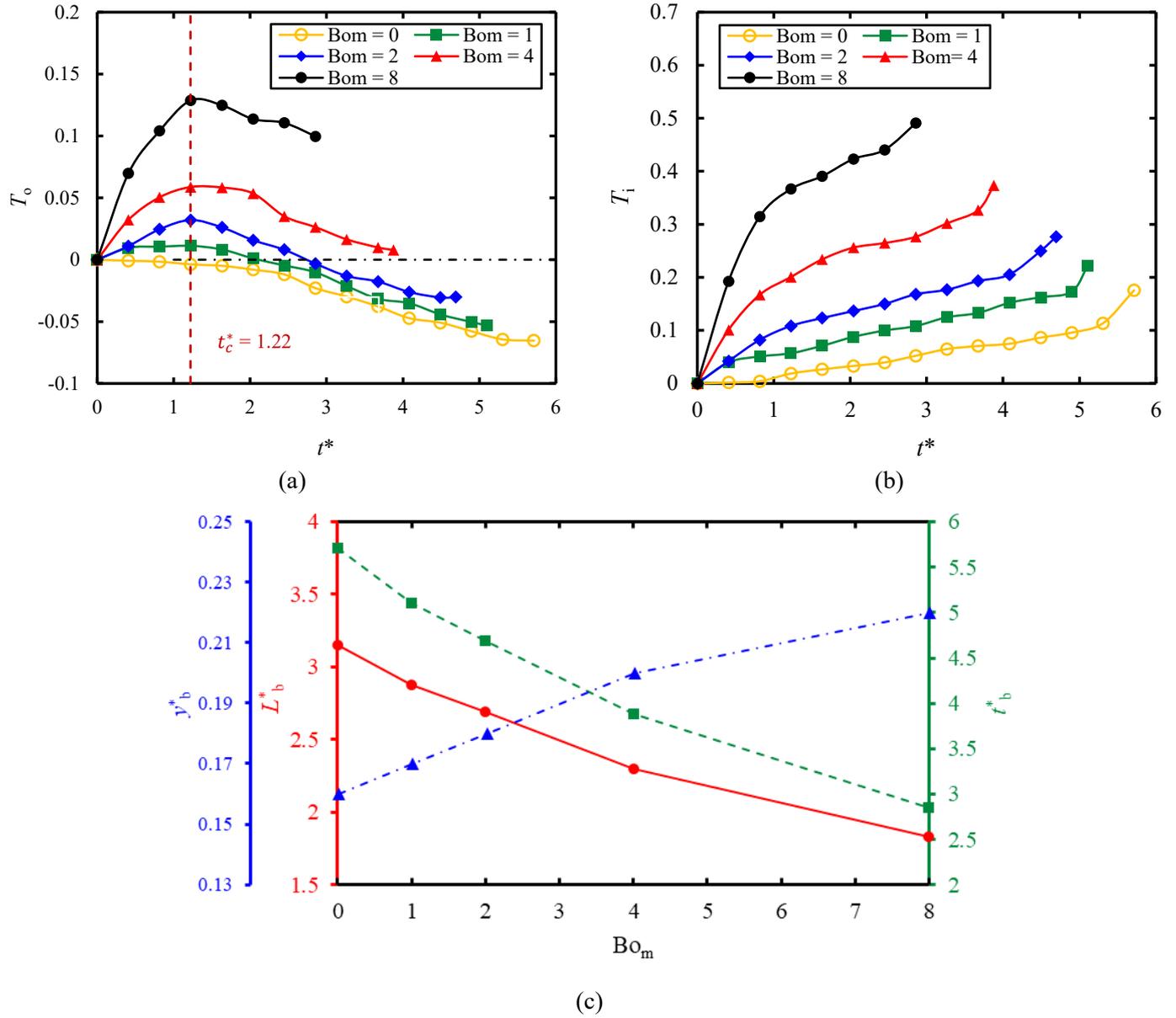

Fig. 7 Influence of the $Bo_m$ number on the FCD deformation and breakup over the course of traversing in a microchannel; (a) variation of the shell deformation with respect to the dimensionless time, (b) variation of the inner droplet deformation with dimensionless time, and (c) $L_b^*$, $t_b^*$, and $y_b^*$ as a function of the magnetic Bond number. The inner droplet is supposed to be ferrofluid and $\alpha = 90°$.

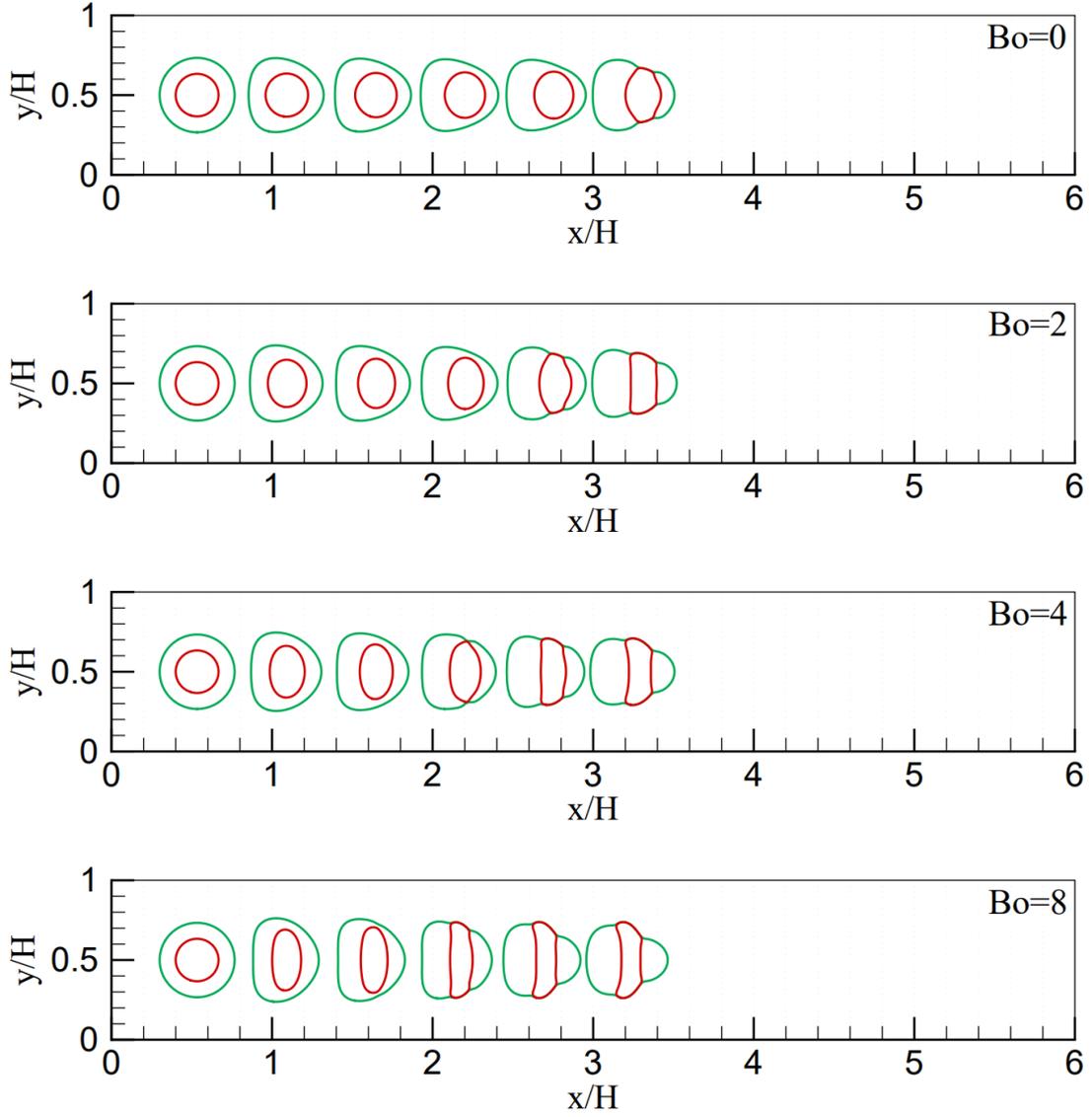

Fig. 8 The migration of an FCD inside a microchannel from the initial point to the outlet over time for various magnetic Bond numbers at $\alpha = 90°$ when the inner droplet is ferrofluid. From left to right: $t^* =$ 0, 1.22, 2.45, 3.67, 4.9 and 6.12.

All in all, selecting the inner droplet as the ferrofluid cannot retard the rupture process for either value of the magnetic field angle. Even when the magnetic field is applied perpendicular to the fluid flow, $t_b^*$ appreciably diminishes. For instance, compared to $Bo_m = 0$, taking the core as ferrofluid and applying a UEMF at $Bo_m = 4$ along $\alpha = 0°$ and $\alpha = 90°$ accelerates the breakup by 5.0% and 32.1%, respectively. Consequently, the shell is presumed to be ferrofluid in section 3.1.2 so that one may figure out whether this choice can delay the rupture phenomenon.

### 3.1.2 The shell is ferrofluid

In the next stage, the influence of the $Bo_m$ number is examined when $\alpha = 0°$ and the shell is ferrofluid ($\chi_2 = 1$), whose results are demonstrated in Fig. 9 and Fig. 10. Concerning Fig. 9, unequivocally, the time variation of inner and outer droplet deformation at Bo = 0 is identical to the results provided in section **Error! Reference source not found.**, where the inner droplet was ferrofluid. Conversely, once the magnetic field is exploited, the FCD morphology notably differs from the former results, which are elaborated on in the following paragraphs.

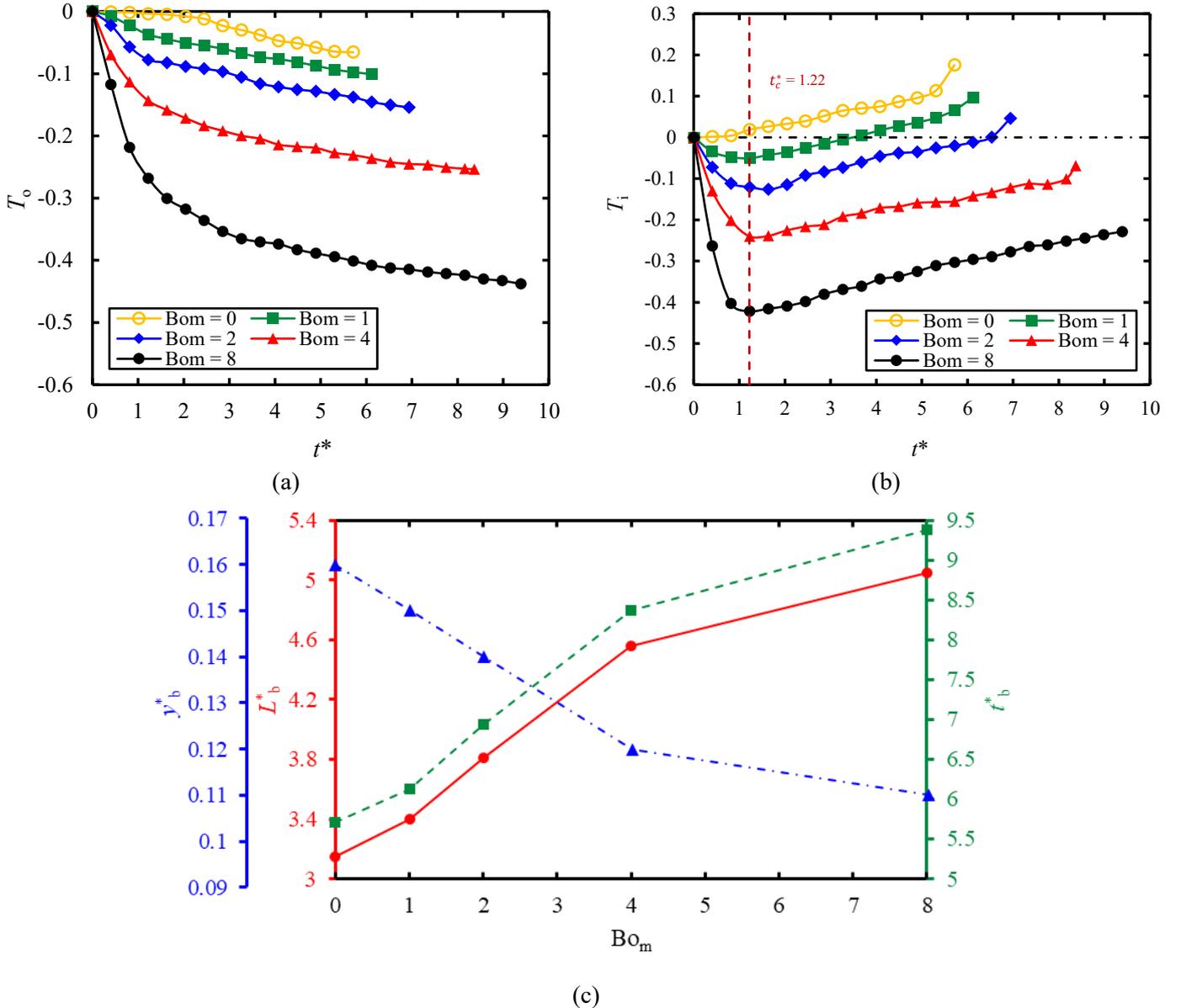

(a)

(b)

(c)

Fig. 9 Influence of the $Bo_m$ number on the FCD deformation and breakup over the course of traversing in a microchannel; (a) variation of the shell deformation with respect to the dimensionless time, (b) variation of the inner droplet deformation with dimensionless time, and (c) $L_b^*$, $t_b^*$, and $y_b^*$ as a function of the magnetic Bond number. The shell is supposed to be ferrofluid and $\alpha = 0°$.

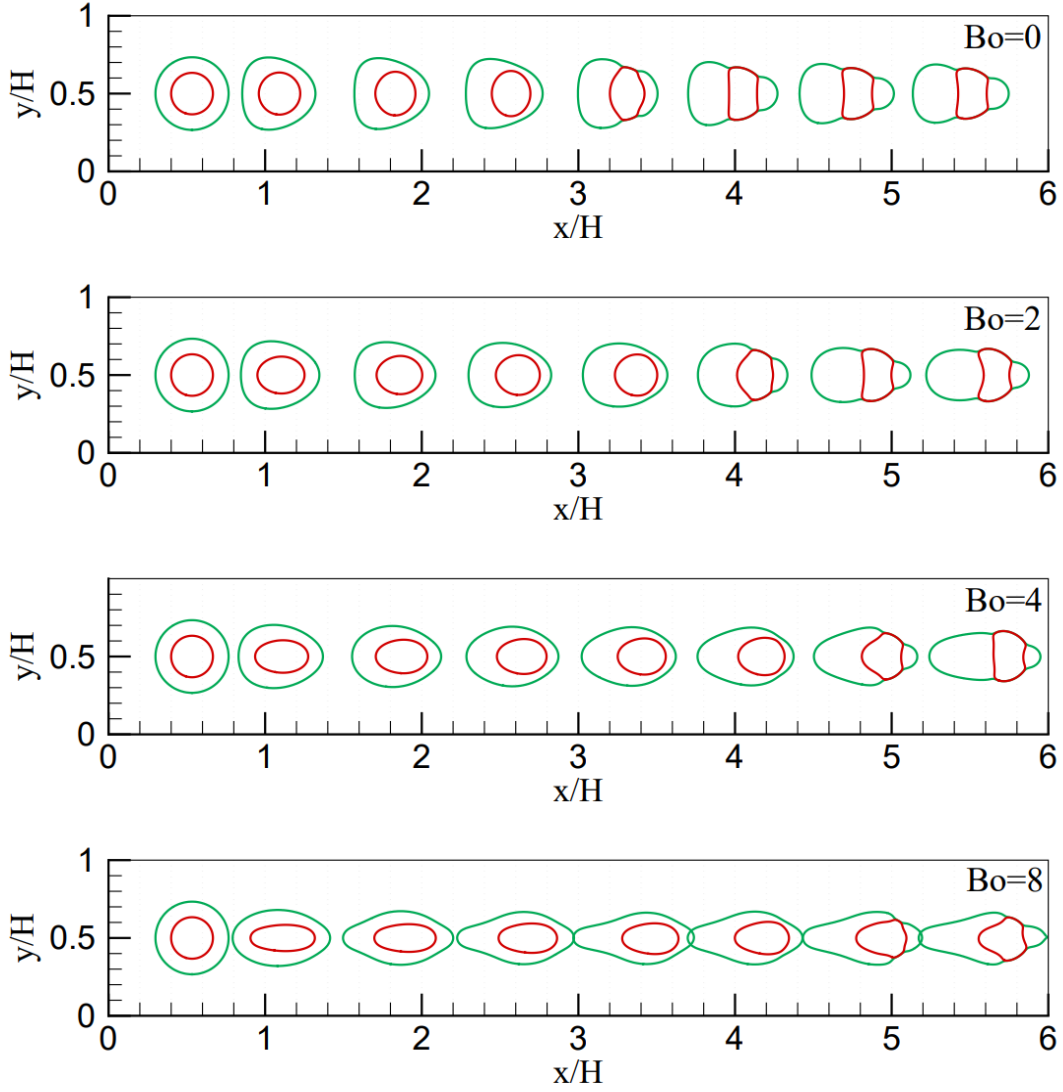

Fig. 10 The migration of an FCD inside a microchannel from the initial point to the outlet over time for various magnetic Bond numbers at $\alpha = 0°$ when the outer droplet is ferrofluid. From left to right: $t^*$ = 0, 1.22, 2.86, 4.49, 6.12, 7.75, 9.39 and 11.02.

It should be stated that when the shell is ferrofluid, the magnetic force is exerted on inner and outer droplets since $\nabla \chi \neq 0$ in both interfaces [see Eq. (7)]. Regarding the outer droplet [Fig. 9 (a)], both the magnetic and drag forces are enforced in the $x$ direction, bringing about an oblate-like configuration, which corresponds to $T_o < 0$ [see Fig. 10]. Additionally, augmenting the Bo number intensifies the magnetic force, which results in a consequential increment in the outer droplet deformation. In the case of the core, the drag force vertically elongates the inner droplet, as substantiated before. On the other hand, since imposed along $\alpha = 0°$, the magnetic force elongates the inner droplet horizontally, giving rise to a competition between these dominant forces, from which the minimum points in

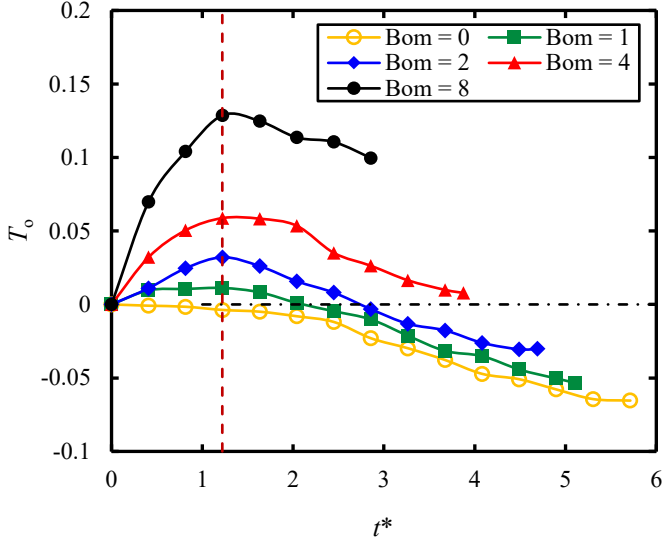
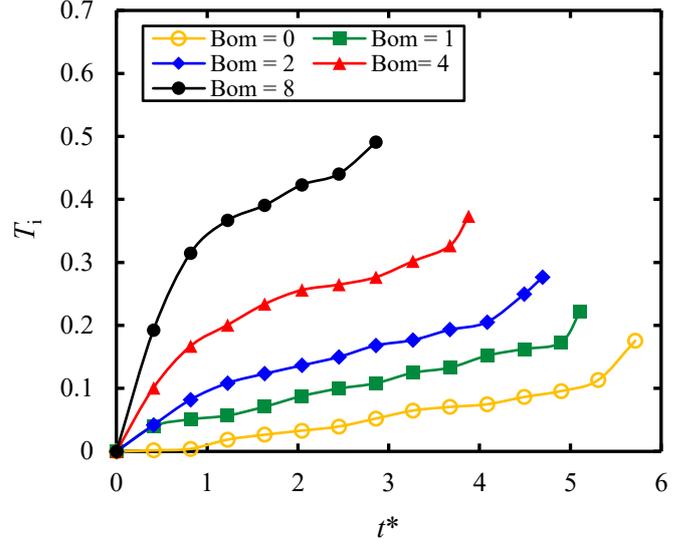

(a)

(b)

(c)

Fig. 7 (b) originate. However, when Bo > 2, the drag force no longer prevails over the magnetic force, conveying that the inner droplet retains the oblate-like shape at the rupture moment, consistent with Fig. 10. It is worth mentioning that the time at which the inner droplet procures the maximum deformation magnitude is the same for various Bo numbers investigated, which is represented by $t_c^*$ in Fig. 9 (b). Also, interestingly, though not ferrofluid, the inner droplet deformation is remarkable.

According to Fig. 9 (c) and Fig. 10, taking the shell to be ferrofluid markedly delays the breakup incidence and raises $L_b^*$ and $t_b^*$ accordingly. To exemplify, compared to $Bo_m = 0$, treating the outer droplet as ferrofluid and exerting a UEMF at $Bo_m = 4$ along $\alpha = 0°$ increases the breakup time by 46.4%. Furthermore, the more strengthened the magnetic field, the later the rupture, complying with the objective of this scholarship. This occurrence may be justified by morphological and hydrodynamics aspects. Morphologically speaking, by augmenting the $Bo_m$ number, the outer droplet elongates from the frontal and rear sides; thus, the inner droplet gains more space to move inside the shell prior to reaching the shell interface, retarding the breakup. Hydrodynamically speaking, since the magnetic force elongates the shell horizontally, prompting an oblate-like shape, the drag force applied to the shell dwindles, which means that it requires a lower force to transport through the ambient flow and can migrate further in advance of the breakup. As expected, $y_b^*$ is reduced with the magnetic Bond number, whose rationale was delineated above.

Fig. 11 and Fig. 12 exhibit the outer and inner droplets' deformation as a function of dimensionless time prior to the rupture and the compound droplet migration process inside the microchannel at $\alpha = 90°$ for various $Bo_m$ numbers, in which the outer droplet is treated as ferrofluid. By virtue of the competition between the magnetic and drag forces, enforced along the *y* and *x* directions, respectively, the outer droplet ends up approaching a bulbous shape by boosting the $Bo_m$ number (see Fig. 12 at $Bo_m = 8$). As anticipated, decreasing the magnetic field intensity reduces the outer droplet deformation, which is lucid in Fig. 11 (a) as well. Concerning Fig. 11 (b), unexpectedly, the inner droplet deformation is more substantial than the shell since not only does the inner droplet face a conspicuous magnetic force despite not being ferrofluid [39], but magnetic and drag forces are exerted in the same direction. These two reasons account for the marked vertical elongation of the inner droplet, drastically accelerating the breakup by raising the $Bo_m$ number, which in turn lowers $t_b^*$ and $L_b^*$ [see Fig. 11 (c)]. Quantitatively speaking, imposing a UEMF at $Bo_m = 4$ and $\alpha = 90°$ expedites the rupture process by 32.1% compared to $Bo_m = 0$. By increasing the $Bo_m$ number, owing to the sharp vertical elongation, the height at which the inner droplet reaches the shell interface rises, bringing about a greater $y_b^*$, as shown in Fig. 11 (c).

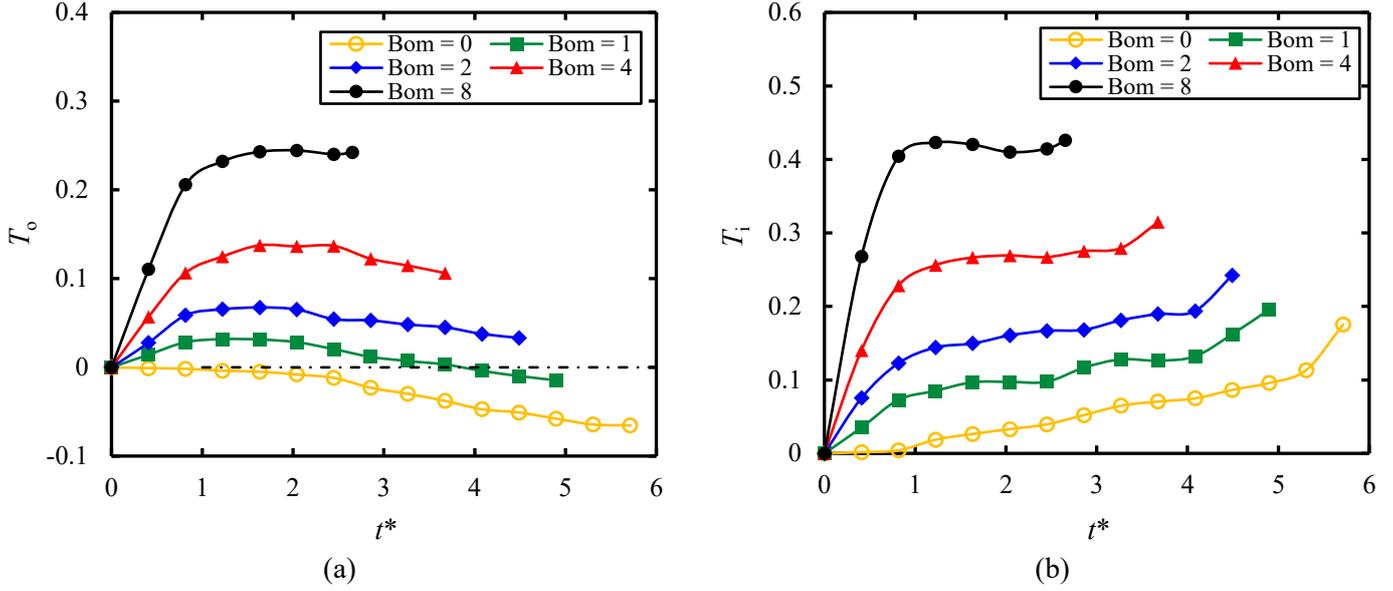

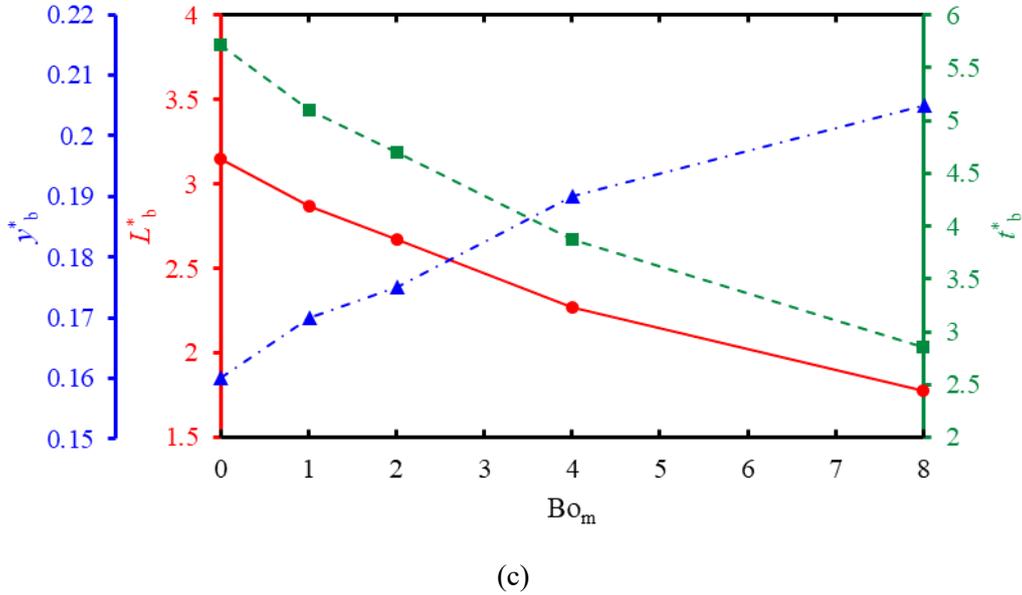

(c)

Fig. 11 Influence of the $Bo_m$ number on the FCD deformation and breakup over the course of traversing in a microchannel; (a) variation of the shell deformation with respect to the dimensionless time, (b) variation of the inner droplet deformation with dimensionless time, and (c) $L_b^*$, $t_b^*$, and $y_b^*$ as a function of the magnetic Bond number. The shell is supposed to be ferrofluid and $\alpha = 90°$.

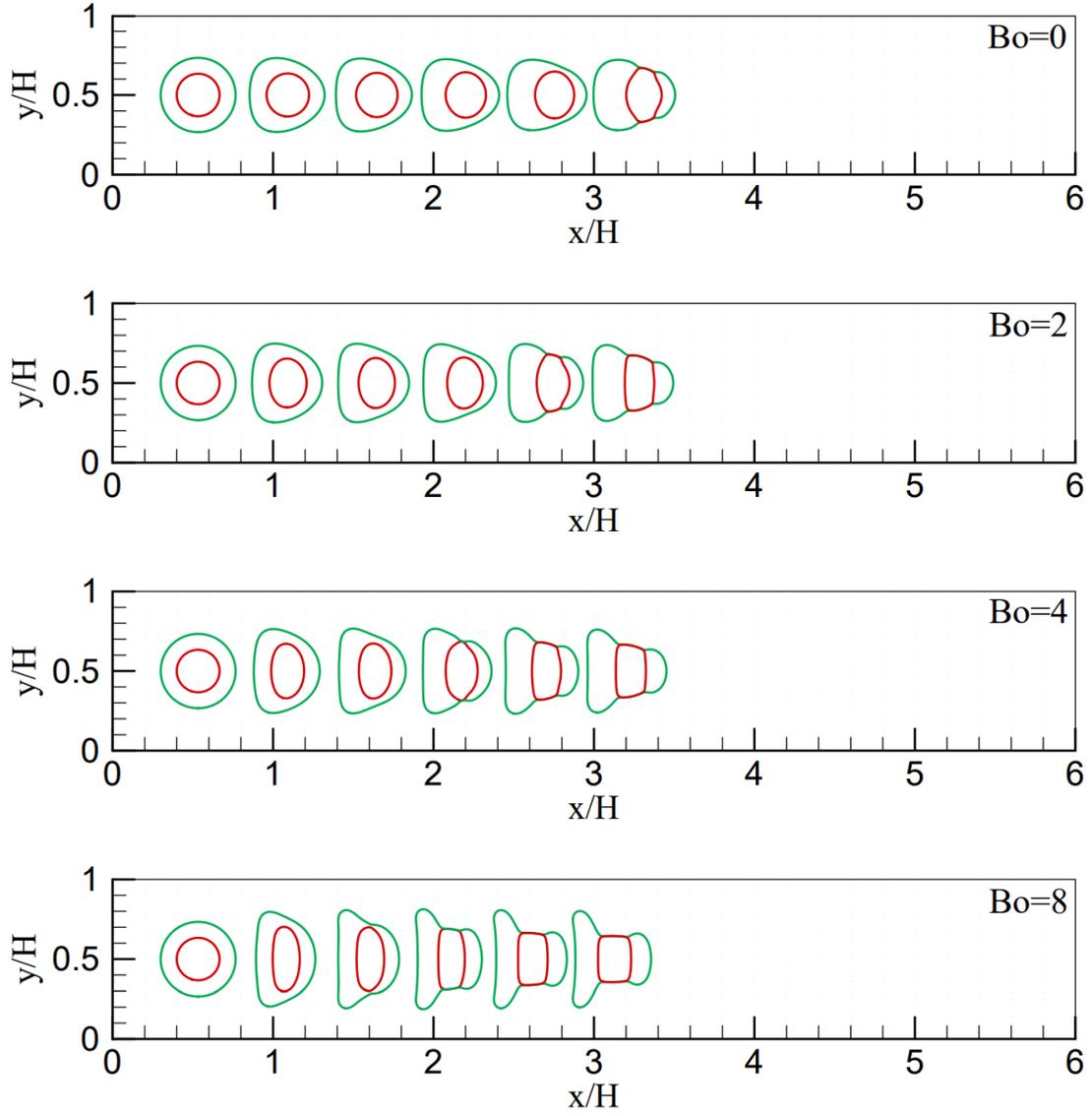

Fig. 12 The migration of an FCD inside a microchannel from the initial point to the outlet over time for various magnetic Bond numbers at $\alpha = 90°$ when the outer droplet is ferrofluid. From left to right: $t^* =$ 0, 1.22, 2.45, 3.67, 4.9 and 6.12.

To better appreciate the reason behind the FCD deformation under the impact of a UEMF, Fig. 13 displays the magnetic force vectors exerted on the FCD along with the pressure field inside the microchannel, in which $p^* = \frac{p}{\rho c_s^2}$ is the normalized pressure. The results are presented at $t^* = 2.45$ and $Bo_m = 2$ for two different values of the magnetic field angles by supposing either the core or shell is ferrofluid. When the core is ferrofluid, it is observed that the magnetic force is not imposed on the outer droplet. Moreover, independent of $\alpha$, a relatively great pressure applied to the shell's rear side generates a flat shape. Additionally, at $\alpha = 0°$, the inner droplet's horizontal elongation increases the pressure inside the shell, causing its frontal side to stretch horizontally, as delineated above. At $\alpha = 90°$, the inner droplet's vertical stretch compresses the fluid in the upper and lower sides of the shell, making it elongate vertically. In contrast, when the shell is ferrofluid, the magnetic force is applied on both interfaces, whose direction is from the ferrofluid to the non-ferrofluid region. Identical to the former case, high pressure on the rear side of the shell results in a flat shape such that at $\alpha = 90°$ the outer droplet approaches a triangular configuration. Also, the pressure inside the inner droplet is noticeably higher than that inside the shell, which is allegedly due to magnetic force direction, pulling the core inward and augmenting the pressure accordingly.

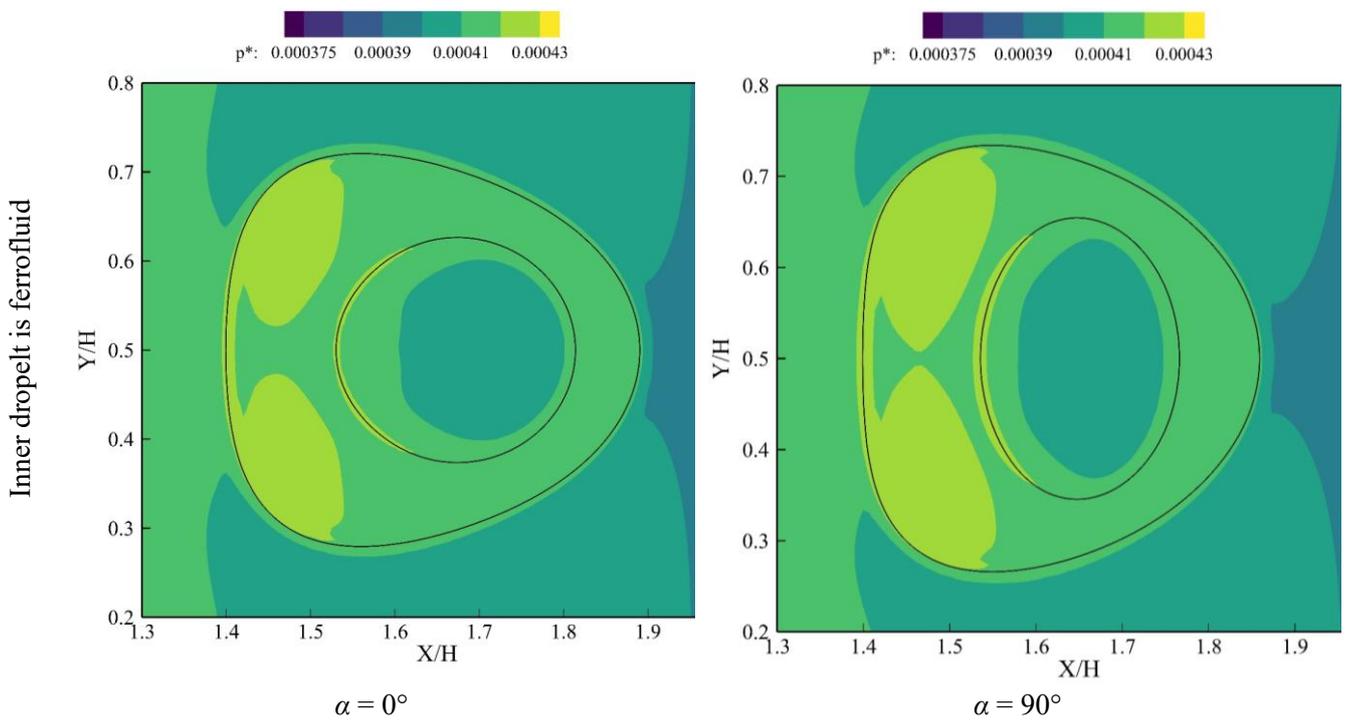

$\alpha = 0°$　　　　　　　　　　　　$\alpha = 90°$

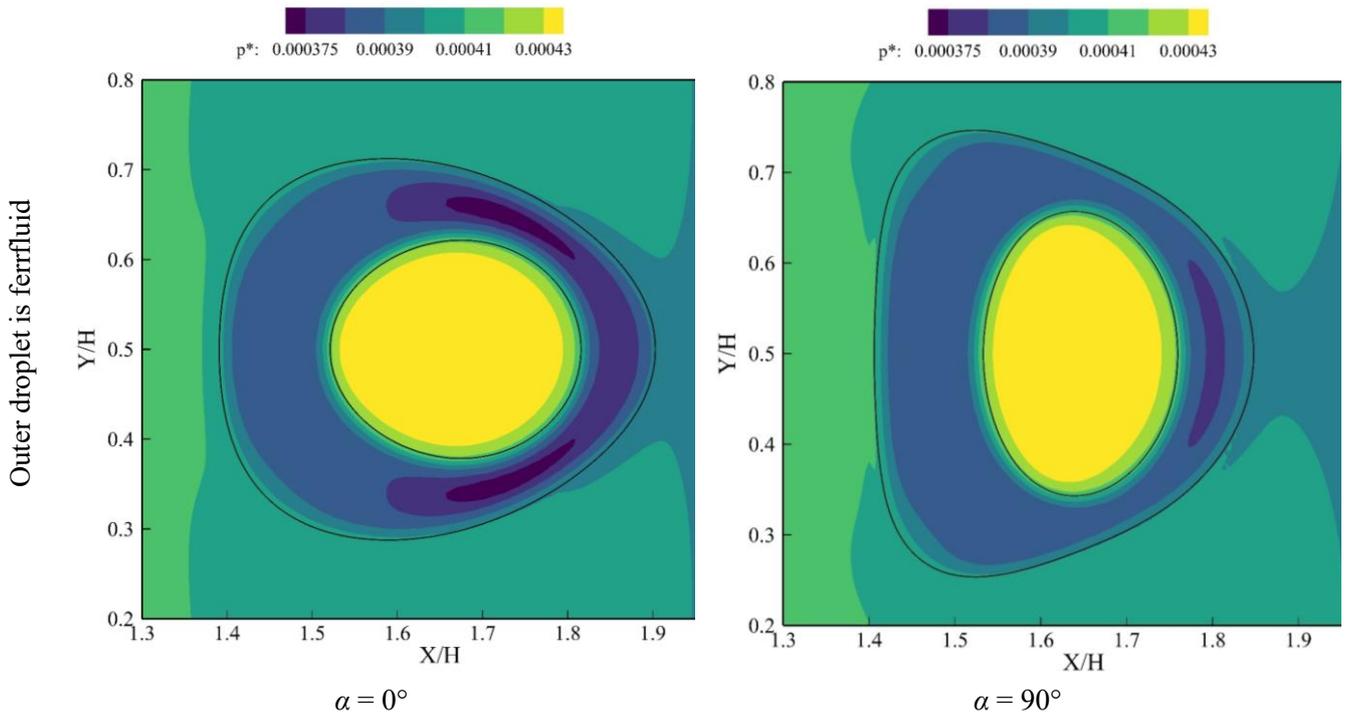

$\alpha = 0°$            $\alpha = 90°$

Fig. 13 Normalized pressure field inside the microchannel and magnetic force vectors applied to an FCD during its migration at $t^* = 2.45$ and $Bo_m = 2$ for different magnetic field angles when either the core or shell is ferrofluid.

To put it shortly, meticulously interpreting the magnetic Bond number impacts, the results yielded in this section proved that taking the shell as ferrofluid and exploiting a UEMF along $\alpha = 0°$ can outstandingly retard the breakup of an FCD moving in a fluid flow. Hence, in the rest of the paper, the shell is assumed to be ferrofluid, and the magnetic field is brought to bear at $\alpha = 0°$ to investigate the effects of other overriding parameters.

## 3.2 Impact of the Capillary number

The Capillary number plays a pivotal role in compound droplet deformation. Microfluidic devices regularly perform in a restricted range of the Ca number. Therefore, in this research, by varying the flow velocity, the Capillary number is altered between 0.05 - 0.30 so that the results are applicable to the pragmatic microfluidic systems. Fig. 14 represents $T_o$ and $T_i$ as a function of time for three distinct values of the Capillary number at $Bo_m = 1$ and $\alpha = 0°$ when the shell is ferrofluid. Fig. 15 exhibits the snapshots of a compound droplet over the course of the movement in the microchannel.

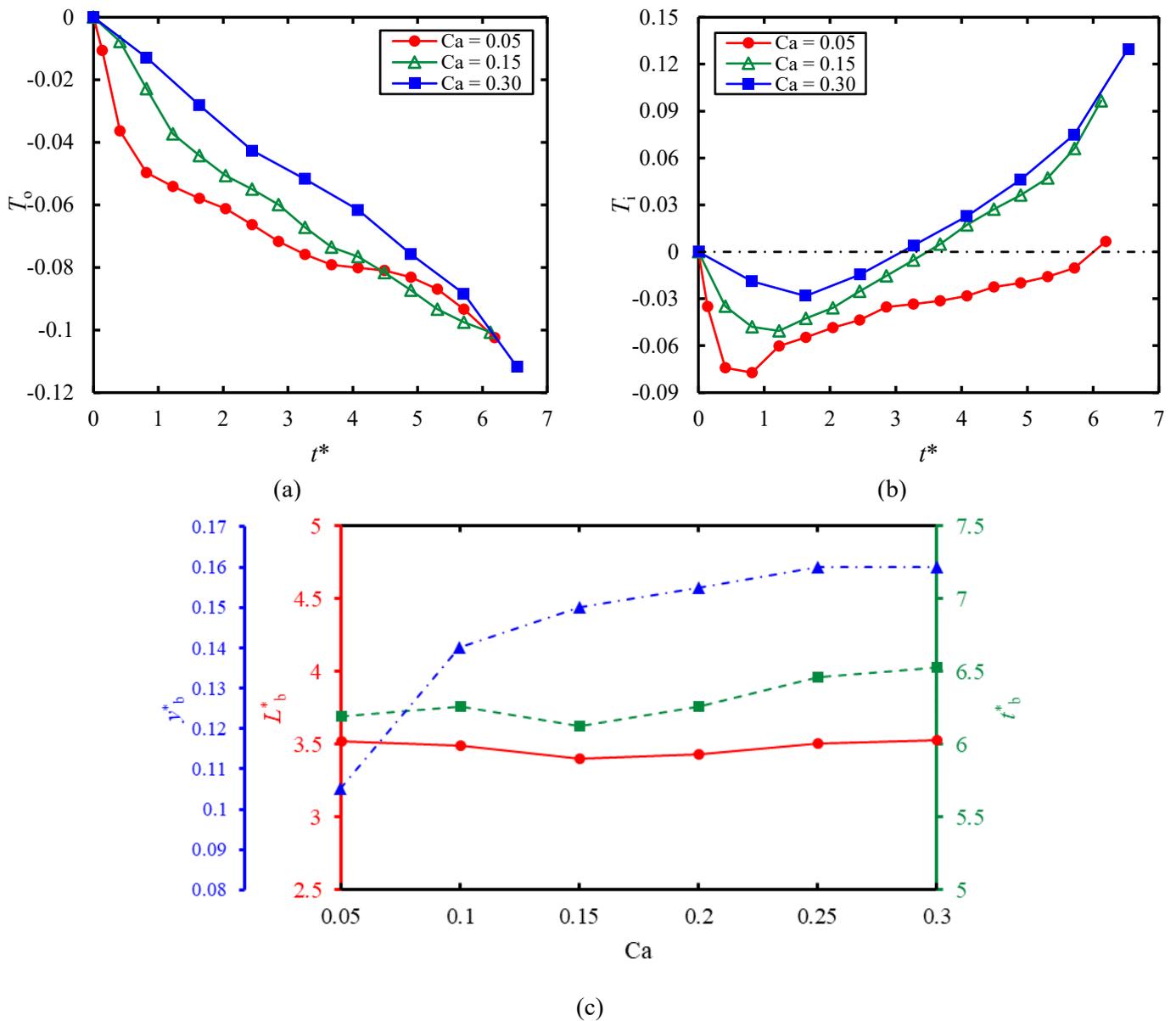

Fig. 14 Influence of the Ca number on the FCD deformation and breakup over the course of traversing in a microchannel at $Bo_m = 1$; (a) variation of the shell deformation with respect to the dimensionless time, (b) variation of the inner droplet deformation with dimensionless time, and (c) $L_b^*$, $t_b^*$, and $y_b^*$ as a function of the Capillary number. The shell is supposed to be ferrofluid and $\alpha = 0°$.

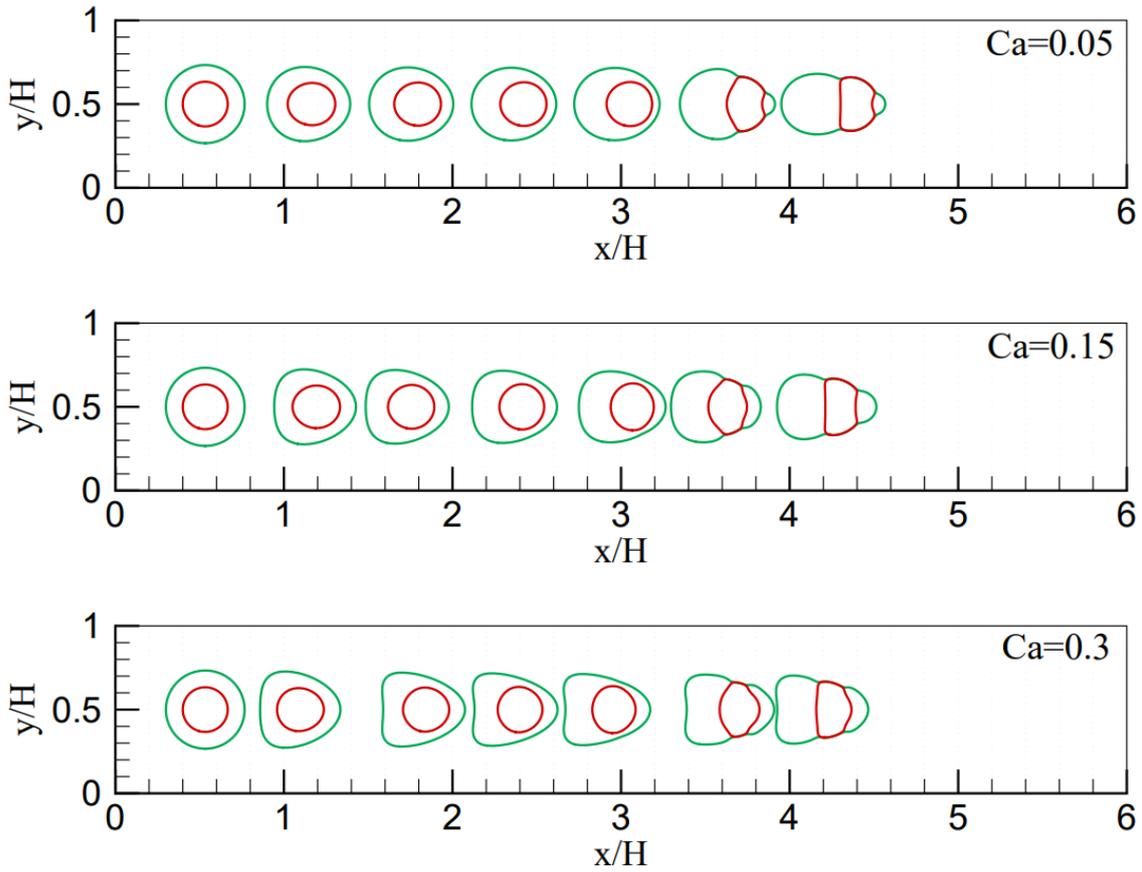

Fig. 15 The migration of an FCD inside a microchannel from the initial point to the outlet over time for three different Capillary numbers at Bo = 1 and $\alpha = 0°$ when the outer droplet is ferrofluid. From left to right: $t^* = 0, 1.36, 2.72, 4.08, 5.44, 6.8$ and $8.16$.

At low Capillary numbers (e.g., Ca = 0.05), the viscous force is inconsequential though the interfacial tension force is notable, which tends to keep the shell in a spherical shape. On the other hand, as it was proven in section 3.1, the magnetic force stretches the rear and frontal sides of the shell concurrently. Consequently, as the viscous force is attenuated, the shell procures an ellipsoid configuration owing to the predominance of the magnetic and interfacial tension forces, which is also evident in Fig. 15 at Ca = 0.05. Furthermore, considering Fig. 14 (a), the smaller the Ca number, the more the shell is horizontally elongated, and hence, the higher the $T_O$ value. By contrast, at the higher values of the Ca number, the interfacial tension force is undermined, while the viscous force is accentuated, which flattens the shell's rear side as described previously. Therefore, at Ca = 0.30, in which the viscous force is conspicuous, a triangular-like shape is attained [see Fig. 15]. Accordingly, the horizontal elongation of the shell's rear side, which results from the magnetic force, is suppressed by the viscous force, which in turn reduces the shell deformation, as pellucid in Fig. 14 (a). Given the inner droplet [Fig. 14 (b)], decreasing the Ca number attenuates the viscous force that competes with magnetic force. Thus, applied horizontally, the magnetic force deforms the inner droplet more straightforwardly, justifying the appreciable deformation of the inner droplet at Ca = 0.05 when $t^* < 1.5$. On the other hand, at $t^* > 1.5$, for the larger values of the Ca number, the viscous force prevails over the magnetic force and eventually deforms the inner droplet into the prolate-like shape.

According to Fig. 14 (c), which plots $y_b^*$, $t_b^*$, and $L_b^*$ as a function of the Ca number, varying the Ca number, caused by flow velocity changes, slightly affects dimensionless rupture length and nondimensional rupture time. Quantitatively speaking, by changing the Ca number from 0.05 to 0.30, $t_b^*$ increases by merely 5.5%, equivalent to a negligible delay in the rupture process. In view of the vertical elongation of the inner droplet with the Ca number increment, it is expected $y_b^*$ grows, which is also axiomatic in Fig. 14 (c).

### 3.3 Impact of the density ratio

This section assesses the effect of the density ratio ($\rho_r$) on the FCD breakup occurrence by altering the shell density while keeping the density of the inner drop and the ambient fluid constant and equal. Five different values of the shell density are considered, corresponding to $1 \leq \rho_r \leq 10$. Fig. 16 depicts the temporal evolution of the inner and outer droplets' deformation in advance of the rupture for various density ratios at $Bo_m = 4$ and $\alpha = 0°$, in which the shell is presumed to be ferrofluid. The transport of the compound droplet in the microchannel over time is revealed in Fig. 17. Contemplating Fig. 16 and Fig. 17, reducing the shell density trivially affects the outer and inner droplets' deformation. Additionally, decreasing $\rho_r$ expedites the rupture process [see Fig. 16 (c)]. Quantitatively speaking, changing the density ratio from 1 to 2 results in a 17.1% decrement in the dimensionless rupture time. Curtailing the density of the outer droplet undermines the viscous drag force the inner droplet experiences [65], assisting it in moving inside the shell more promptly. Therefore, the inner droplet touches the interface of the shell more swiftly, making the breakup occur sooner, which is also clear in Fig. 17. Based on Fig. 16 (c), as the density ratio grows, $y_b^*$ declines since lessening the outer droplet density (increasing $\rho_r$), alleviates the viscous drag force exerted on the core, decreasing its vertical elongation, which in turn diminishes the height at which the inner droplet reaches the interface of the shell.

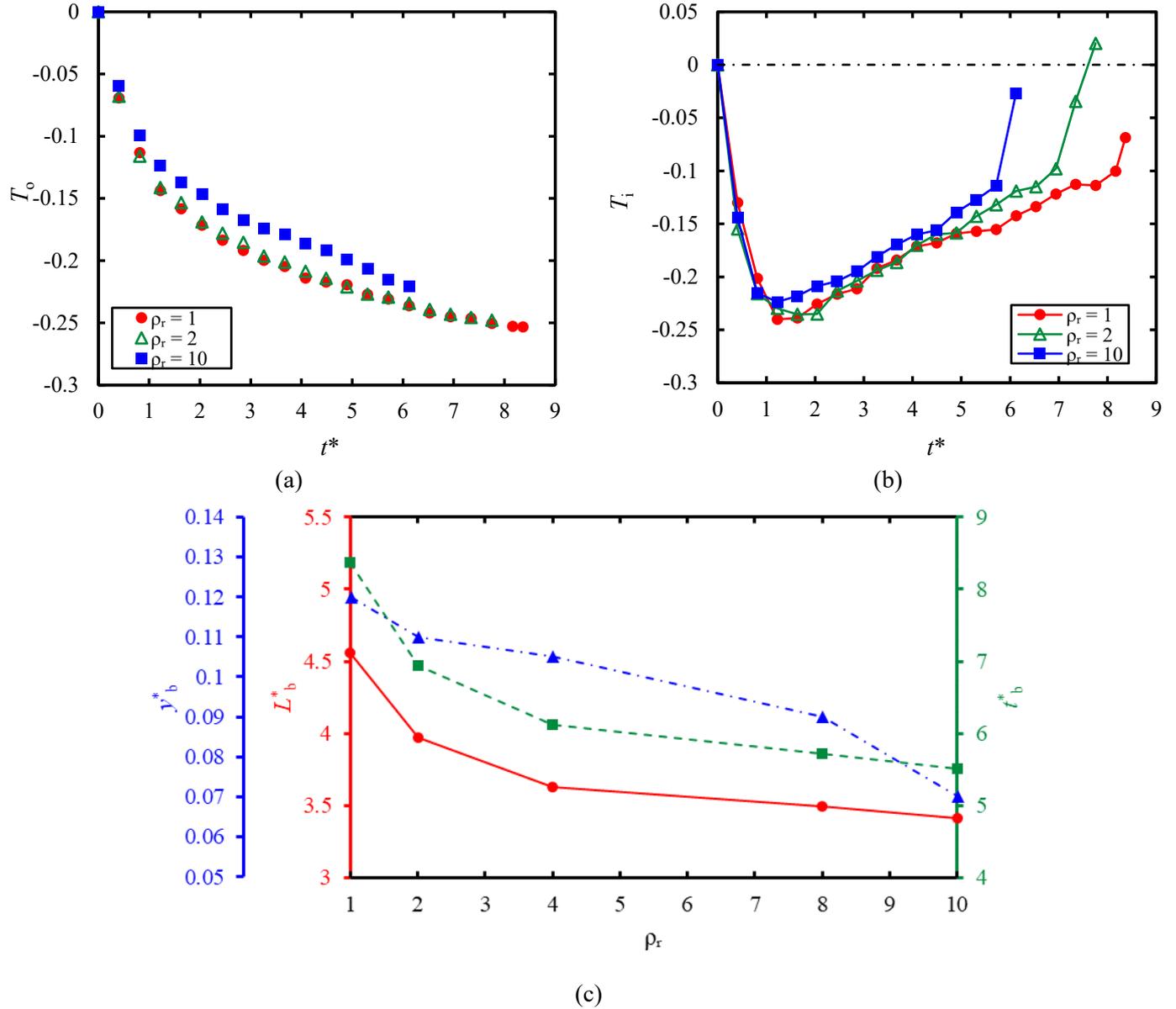

Fig. 16 Influence of the density ratio ($\rho_r$) on the FDC deformation and breakup over the course of traversing in a microchannel at $Bo_m = 4$; (a) variation of the shell deformation with respect to the dimensionless time, (b) variation of the inner droplet deformation with dimensionless time, and (c) $L_b^*$, $t_b^*$, and $y_b^*$ as a function of the density ratio. The shell is supposed to be ferrofluid and $\alpha = 0°$.

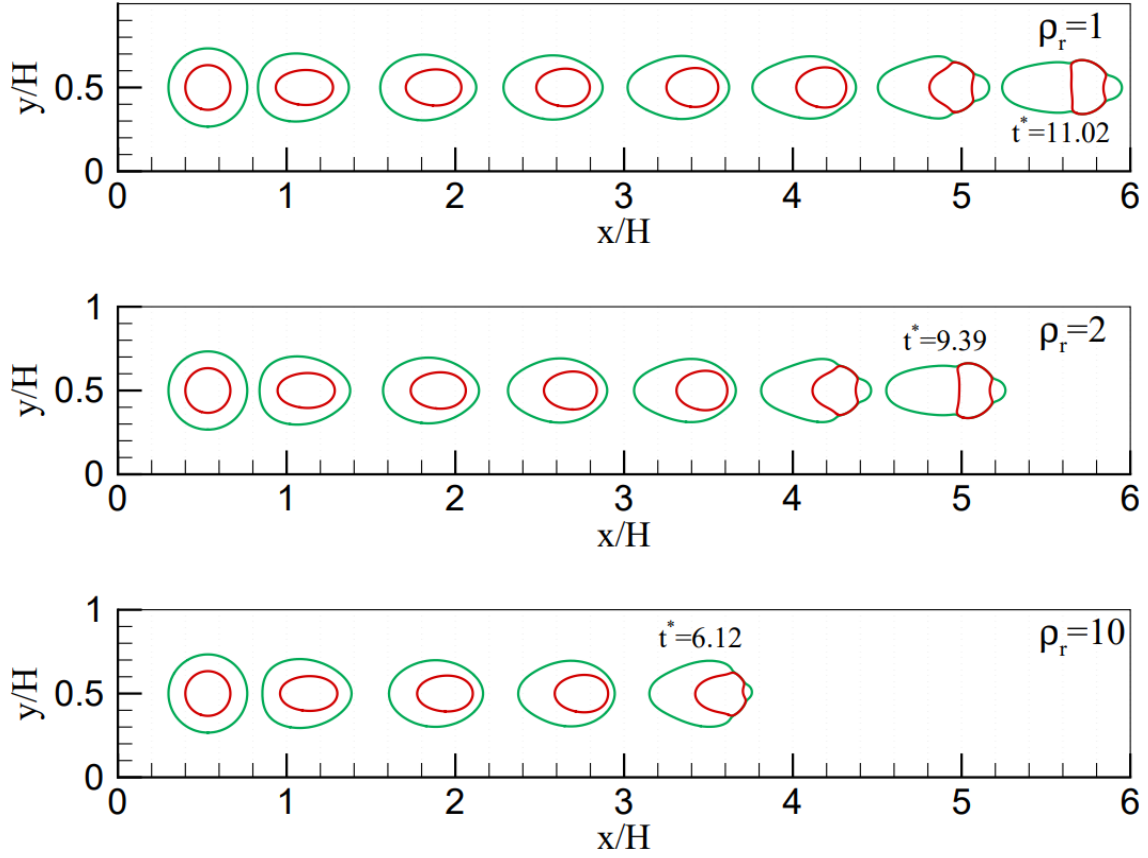

Fig. 17 The migration of an FCD inside a microchannel from the initial point to the outlet over time for three different density ratios at $Bo_m = 4$ and $\alpha = 0°$ when the outer droplet is ferrofluid. From left to right: $t^* = 0, 1.22, 2.86, 4.49, 6.12, 7.75, 9.39$ and $11.02$.

### 3.4 Impact of the viscosity ratio

Changing the shell viscosity while maintaining the viscosity of the inner droplet and ambient fluid constant and equal, this section scrutinizes the viscosity ratio ($v_r$) effects. Five distinct values of the shell viscosity are picked out, equivalent to $1 \leq v_r \leq 10$. Hence, the ternary system operates as an O/W/O configuration. Fig. 18 and Fig. 19 show the time variation of the FCD deformation and the snapshots of the FCD transport inside the microchannel for different viscosity ratios at $Bo_m = 1$ and $\alpha = 0°$, in which the shell is assumed to be ferrofluid. The greater the viscosity, the closer the shell behavior to the solid, and in conclusion, the less deformed the shell, which elucidates the reason for the decline in outer droplet deformation as the viscosity ratio diminishes [see Fig. 18 (a) and Fig. 19]. Interestingly, the influence of the viscosity ratio on the compound droplet deformation under a UEMF is similar to that in the absence of the magnetic field, which is reported by Liu et al. [18] and Vu et al. [66]. Moreover, by augmenting the viscosity ratio (reducing the shell viscosity), the magnetic force may deform the inner droplet more simply when $t^* < 1.22$, causing it to be more deformed horizontally [see Fig. 18 (b) and Fig. 19]. Conversely, by lowering the viscosity ratio, in the wake of the higher viscous drag force, the inner droplet ends up conspicuously deforming vertically at the rupture moment. Given Fig. 18 (c), raising the viscosity ratio precipitates the breakup process. To put it differently, for the lowest value of the viscosity ratio ($v_r = 1$), the FCD can further transport in the microchannel prior to encountering the rupture. Indeed, boosting the shell viscosity (decreasing the viscosity ratio) raises the force the inner droplet requires to move forward and contact the shell interface, which retards the breakup. Last but not least, regarding Fig. 18 (c), $y_b^*$ is not altered by the viscosity ratio.

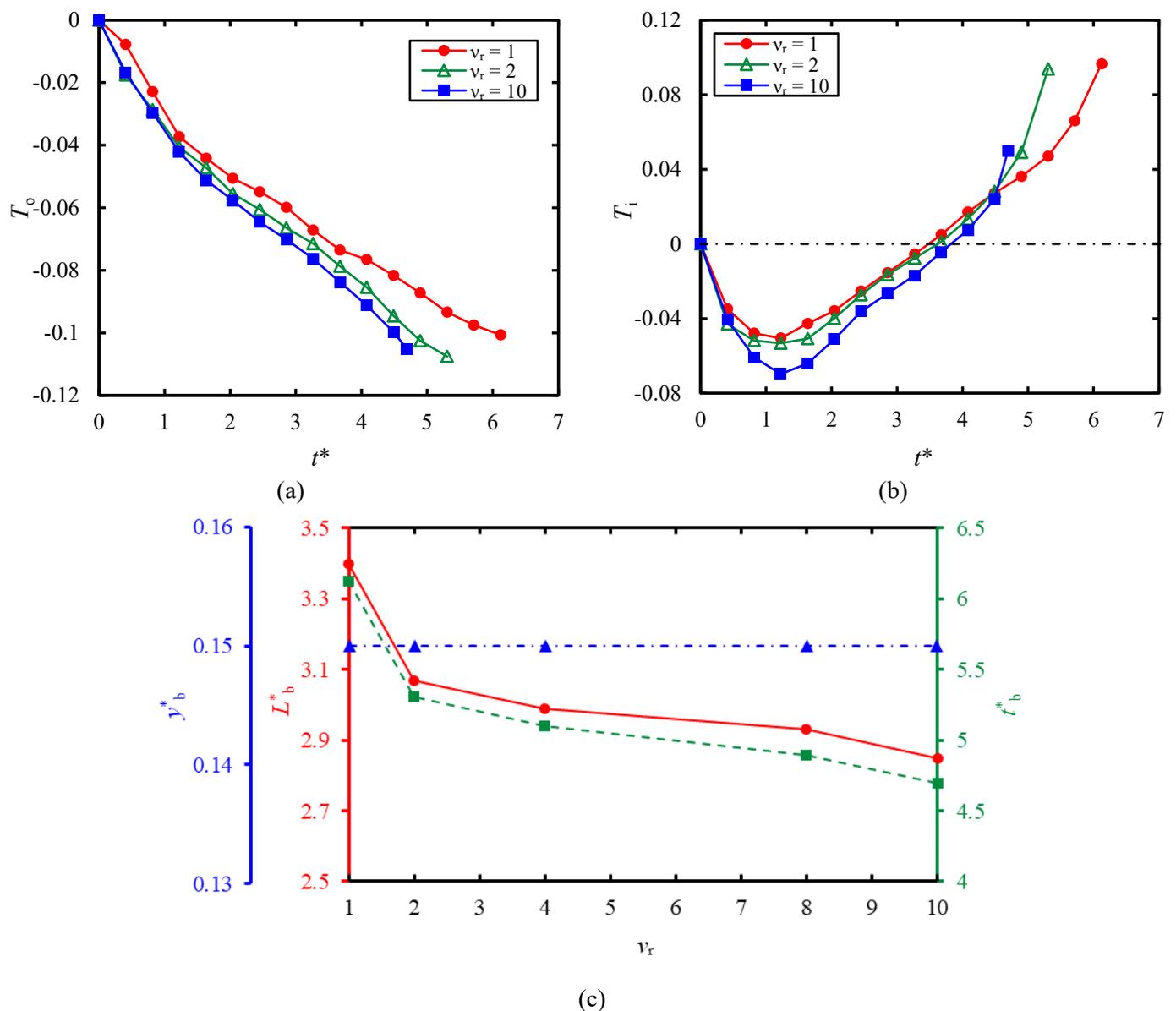

(a)

(b)

(c)

Fig. 18 Influence of the viscosity ratio ($v_r$) on the FCD deformation and breakup over the course of traversing in a microchannel at $Bo_m = 1$; (a) variation of the shell deformation with respect to the dimensionless time, (b) variation of the inner droplet deformation with dimensionless time, and (c) $L_b^*$, $t_b^*$, and $y_b^*$ as a function of the viscosity ratio. The shell is supposed to be ferrofluid and $\alpha = 0°$.

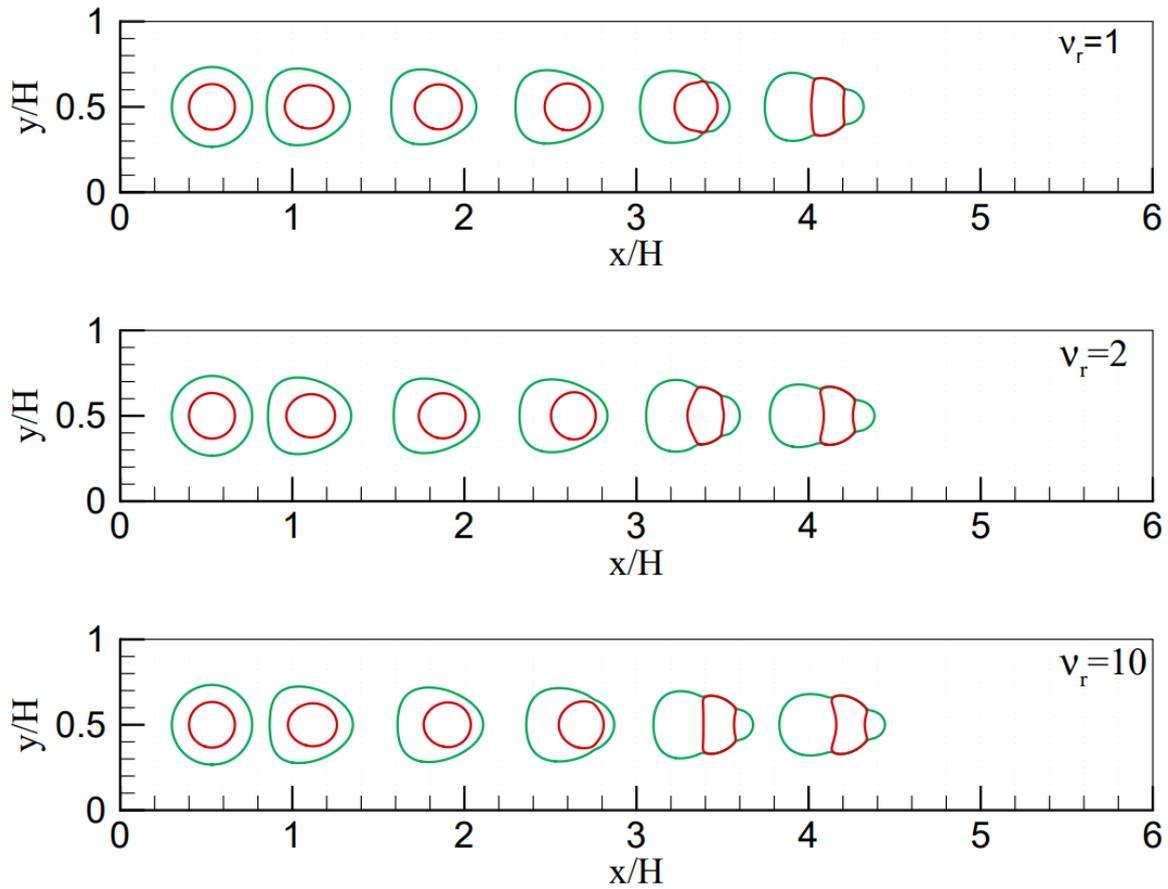

Fig. 19 The migration of an FCD inside a microchannel from the initial point to the outlet over time for three different viscosity ratios at $Bo_m = 1$ and $\alpha = 0°$ when the outer droplet is ferrofluid. From left to right: $t^* = 0$, 1.22, 2.86, 4.49, 6.12 and 7.75.

### 3.5  Impact of the radius ratio

Preserving the outer droplet radius at 35 lattice units and altering the inner droplet radius, this section is allotted to surveying the influence of the radius ratio ($r_r$). The considered radii for the inner droplet in lattice units are 15, 20, and 25, which correspond to $r_r = 2.3$, $r_r = 1.75$, and $r_r = 1.4$, respectively. The changes in inner and outer droplets' deformation with dimensionless time before the rupture for three distinct radius ratios are depicted in Fig. 20 at $Bo_m = 2$ and $\alpha = 0°$ when the shell is ferrofluid. Also, Fig. 21 provides the traversing process of the FCD. The shell space is confined as the radius ratio declines. Accordingly, by a finite deformation, the inner droplet touches the shell interface, provoking the breakup. In other words, reducing the radius ratio dramatically accelerates the breakup process, which is consistent with Fig. 20 and Fig. 21. The swift rupture prompted by the inner droplet's sizeable radius (low radius ratio) prevents the outer droplet from being exposed to the magnetic and drag forces for a sufficient time prior to the rupture, exceedingly limiting its deformation, which is obvious in Fig. 20 (a). Intriguingly, when $t^* < 1.0$, $T_o$ does not rely on the radius ratio and solely depends on time. According to Fig. 20 (b), by increasing the core radius (curtailing the radius ratio), the inner droplet undergoes a more substantial horizontal deformation when $t^* < 1.5$. That is because enlarging the core size impedes the flow development inside the shell [20], diminishing the velocity magnitude in the shell, which alleviates the drag force applied to the inner droplet. Thus, the magnetic force can further deform the core horizontally. Additionally, the decrement in the drag force allows the core to transport inside the shell more promptly, markedly precipitating the rupture process [see Fig. 20 (c)]. To exemplify, decreasing the radius ratio from 2.3 to 1.75 accelerates the rupture incidence by 23.9%. Distinctly, a decrement in the inner droplet size (raising $r_r$) lessens $y_b^*$.

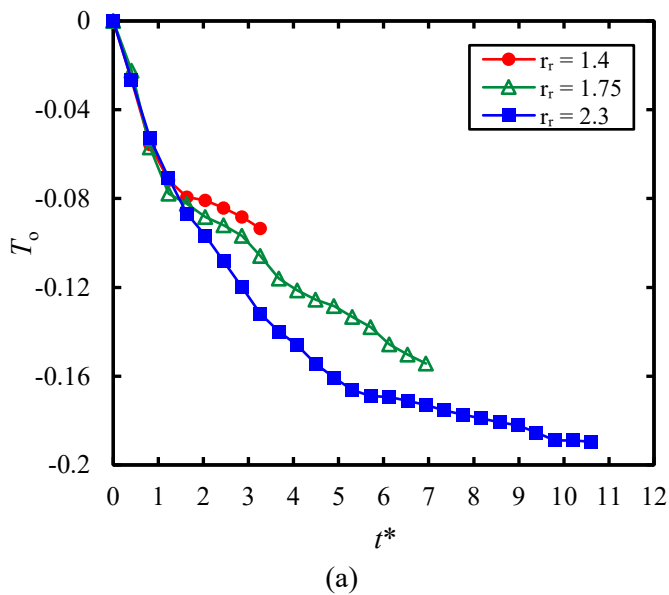
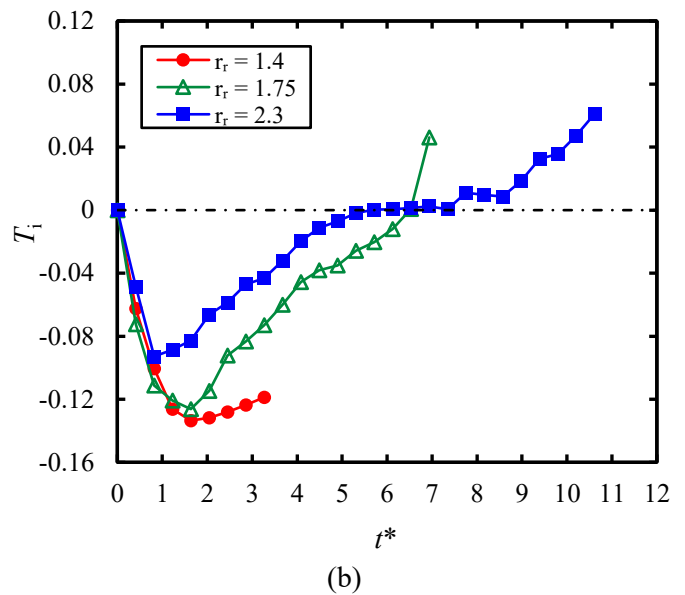

(a)          (b)

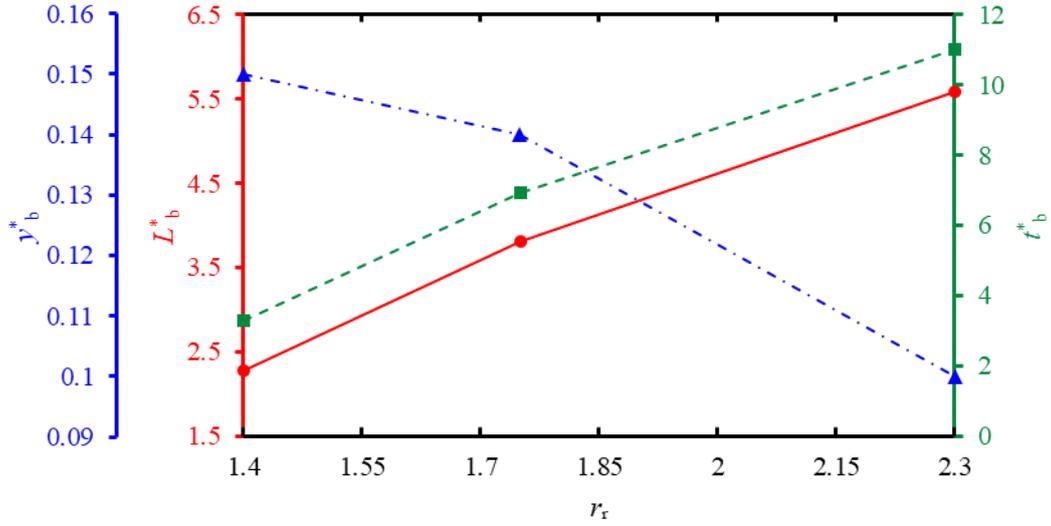

(c)

Fig. 20 Influence of the radius ratio ($r_r$) on the FCD deformation and breakup over the course of traversing in a microchannel at $Bo_m = 2$; (a) variation of the shell deformation with respect to the dimensionless time, (b) variation of the inner droplet deformation with dimensionless time, and (c) $L_b^*$, $t_b^*$, and $y_b^*$ as a function of the radius ratio. The shell is supposed to be ferrofluid and $\alpha = 0°$.

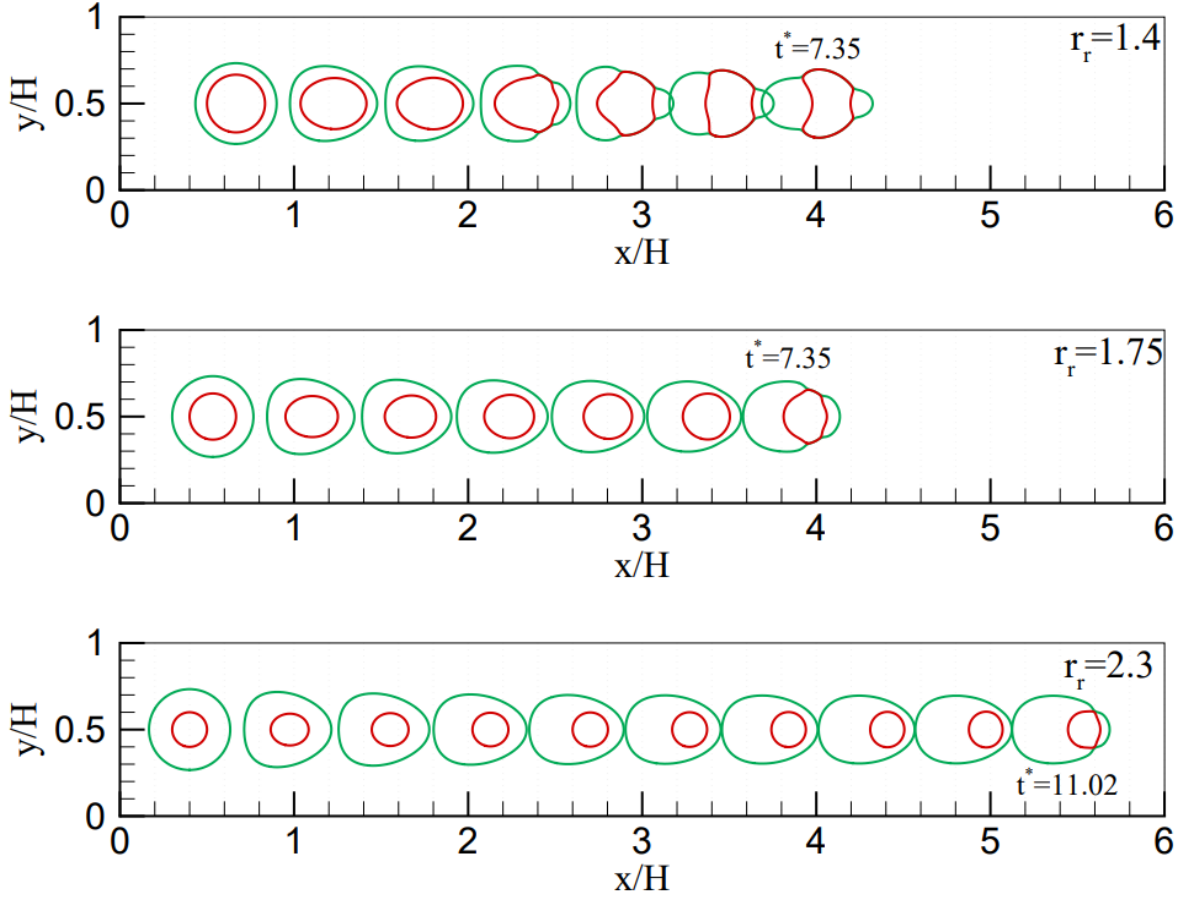

Fig. 21 The migration of an FCD inside a microchannel from the initial point to the outlet over time for three different radius ratios at $Bo_m = 2$ and $\alpha = 0°$ when the outer droplet is ferrofluid. From left to right: $t^* = 0, 1.22, 2.45, 3.67, 4.9, 6.12, 7.35, 8.57, 9.79$ and $11.02$.

### 3.6 Impact of the surface tension coefficients

Surface tension between the involved phases critically impacts the FCD morphology. To explore the effects of surface tension coefficients, disparate cases are considered utilizing the stability diagram provided by Guzowski et al. [67], where the feasible equilibrium topologies of two immiscible droplets immersed in a third fluid are proposed. The examined topologies are defined below:

1) $\left|\frac{\sigma_{13}}{\sigma_{12}} - \frac{\sigma_{23}}{\sigma_{12}}\right| < 1$ and $\frac{\sigma_{13}}{\sigma_{12}} + \frac{\sigma_{23}}{\sigma_{12}} > 1$, where the droplets of the first and second phases bear a common interface and are subjected to the third phase, which is called partial engulfing.

2) $\frac{\sigma_{13}}{\sigma_{12}} + \frac{\sigma_{23}}{\sigma_{12}} < 1$, where the pair of simple droplets remains separated [12], which is called non-engulfing.

3) $\frac{\sigma_{13}}{\sigma_{12}} > \frac{\sigma_{23}}{\sigma_{12}} + 1$, where the first phase is engulfed utterly by the droplet of the second phase, which is called complete engulfing-type 1, under which the core is kept inside the shell for a noticeable time.

4) $\frac{\sigma_{23}}{\sigma_{12}} > \frac{\sigma_{13}}{\sigma_{12}} + 1$, where the second phase is engulfed utterly by the droplet of the first phase, which is called complete engulfing-type 2.

Table 2 summarizes the ratio of surface tension coefficients for the various cases discussed above. It should be expressed that $\sigma_{23}$ is kept constant in all examined cases so that the Capillary number is not changed [see Eq. (37)].

Table 2 Surface tension ratios of various equilibrium topologies considered in this research.

| Case | $\sigma_{23}/\sigma_{12}$ | $\sigma_{13}/\sigma_{12}$ |
|---|---|---|
| non-engulfing | 0.30 | 0.30 |
| complete engulfing-type 1 | 0.50 | 1.70 |
| partial engulfing | 1.0 | 1.0 |
| complete engulfing-type 2 | 1.70 | 0.50 |

Fig. 22 illustrates the time evolution of the deformation of the outer and inner droplets prior to the breakup for different surface tension coefficients at Bo$_m$ = 4 and $\alpha = 0°$ when the shell is treated as ferrofluid. Interestingly, the core deformation has a direct relation with the $\sigma_{23}/\sigma_{12}$ ratio, while the outer droplet deformation has an inverse relation with this parameter. Since $\sigma_{23}$ remains constant, $T_o$ and $T_i$ actually vary with the surface tension coefficient between the inner and outer droplet ($\sigma_{12}$). Thus, though boosting the shell deformation, increasing $\sigma_{12}$ diminishes the inner droplet deformation. That is expected as the interfacial tension force impedes the core deformation. This elucidation concurs with Fig. 22, in which the non-engulfing case bearing the lowest amount of $\sigma_{23}/\sigma_{12}$ instigates large deformation of the outer droplet while hindering the significant deformation of the inner droplet. By contrast, bearing the highest value of $\sigma_{23}/\sigma_{12}$, the complete engulfing-type 2 case suppresses outer droplet deformation though causing a remarkable deformation of the inner droplet. This occurrence is clear-cut in Fig. 23 as well where the transport process of the FCD is shown for several surface tension coefficients. It should be mentioned that in advance of the rupture, inasmuch as the inner droplet and ambient fluid are not in direct contact, $T_i$ does not rely on $\sigma_{13}$. However, $\sigma_{13}$ plays a crucial role after the breakup as the inner droplet is exposed to the ambient fluid. Concerning Fig. 23, in the case of non-engulfing, rupture occurs expeditiously since the frontal side of the shell does not maintain the spherical shape and becomes narrower gradually. Therefore, a nose-like structure is eventually created, which restricts the region between the inner and outer droplets, provoking the breakup in a short time ($t_b^* = 5.10$).

Conversely, the complete engulfing-type 1 condition noticeably delays the rupture incidence as predicted. Dimensionless rupture time ($t_b^*$), dimensionless rupture length ($L_b^*$), and nondimensional rupture height with respect to the channel center line ($y_b^*$) are listed in Table 3. Contemplating Table 3, one may realize that by altering the surface tension coefficient so that the complete engulfing-type 1 condition is procured, $t_b^*$ and $L_b^*$ grows by 34.2% and 29.6%, respectively, compared to the case at which all surface tension coefficients are equal (partial engulfing), which is equivalent to a sharp delay in the breakup. On the other hand, surface tension coefficients inconsequentially affect $y_b^*$.

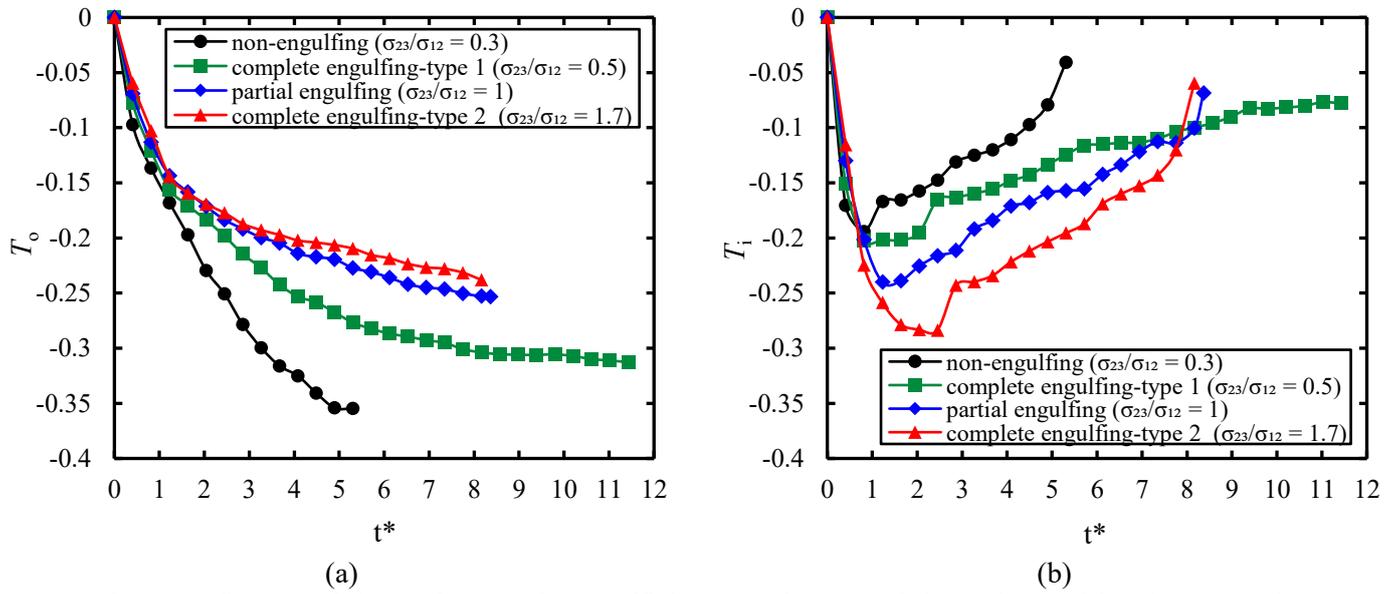

(a)          (b)

Fig. 22 Influence of the surface tension coefficients on the FCD deformation and breakup over the course of traversing in a microchannel at $Bo_m = 4$; (a) variation of the shell deformation with respect to the dimensionless time, (b) variation of the inner droplet deformation with dimensionless time. The shell is supposed to be ferrofluid and $\alpha = 0°$.

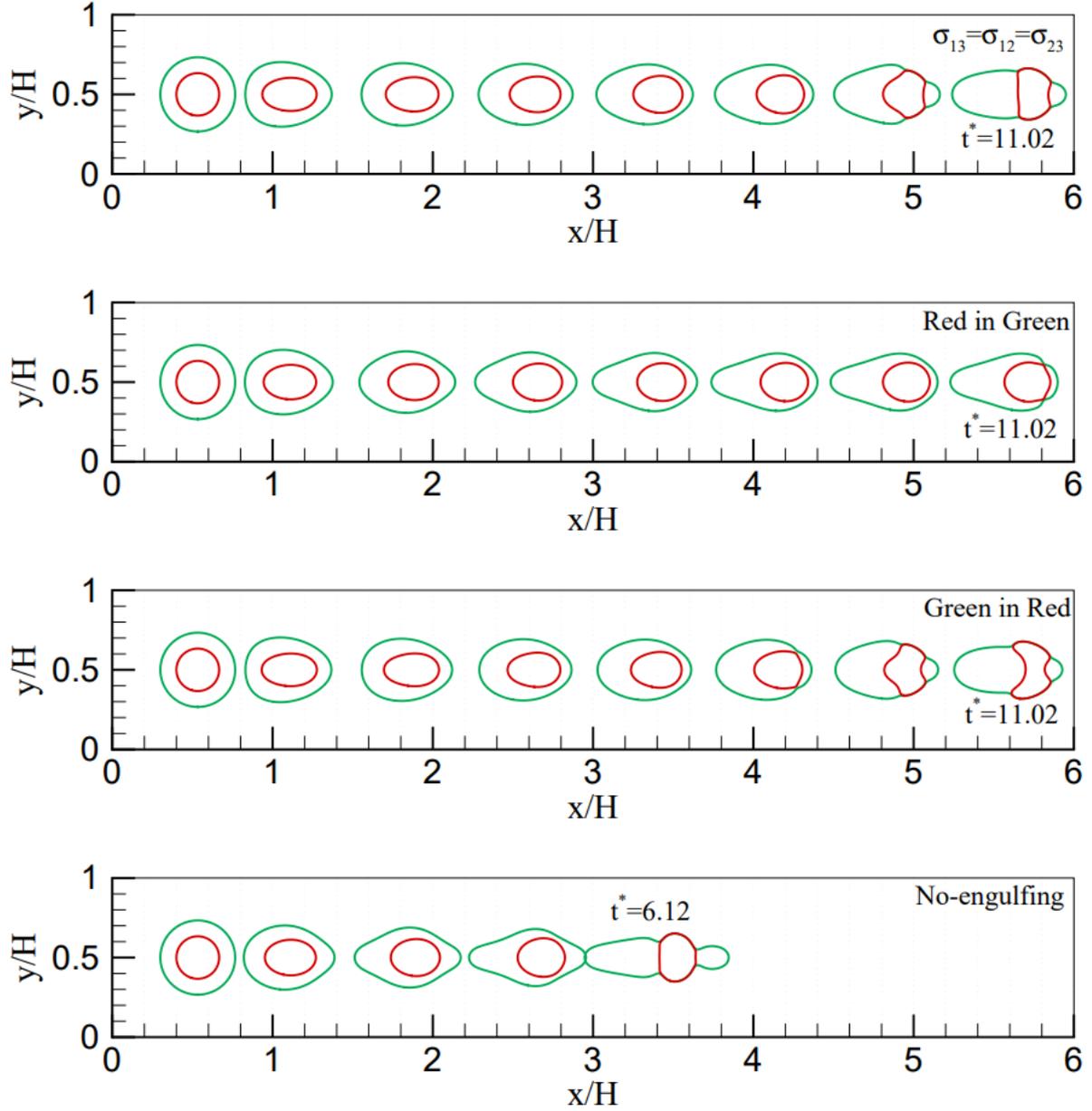

Fig. 23 The migration of an FCD inside a microchannel from the initial point to the outlet over time for disparate surface tension coefficients at $Bo_m = 4$ and $\alpha = 0°$ when the outer droplet is ferrofluid. From left to right: $t^* = 0, 1.22, 2.86, 4.49, 6.12, 7.75, 9.39$ and $11.02$.

Table 3 $L_b^*$, $t_b^*$, and $y_b^*$ for various equilibrium topologies considered in this research at $Bo_m = 4$ and $\alpha = 0°$. The shell is presumed to be ferrofluid.

| Case | $L_b^*$ | $t_b^*$ | $y_b^*$ |
|---|---|---|---|
| non-engulfing | 3.06 | 5.10 | 0.115 |
| complete engulfing-type 1 | 5.91 | 11.22 | 0.11 |
| partial engulfing | 4.56 | 8.37 | 0.12 |
| complete engulfing-type 2 | 4.46 | 8.16 | 0.12 |

## 3.7 Regime diagrams of the deformation and breakup

In this section, the morphology of the FCD is analyzed thoroughly when the shell is ferrofluid and the UEMF is applied along $\alpha = 0°$, the circumstance under which the onset of the breakup is strikingly postponed (see section 3.1). By varying the $Bo_m$ and Ca numbers, five distinguished regimes were observed: semi-spherical, triangular-like, egg-like-type 1, egg-like-type 2, and tadpole-like, all of which are discussed below.

Fig. 24 displays a typical example of the observed regimes. Depicted in Fig. 24 (a), the semi-spherical regime happens at the small values of magnetic Bond ($Bo_m \leq 2$) and Capillary ($Ca \leq 0.15$) numbers. In light of the gentle magnetic force, the outer droplet elongation is small (e.g., for the shown case, $T_o = 0.047$ at the rupture moment). Moreover, due to the small Ca number, the interfacial tension force is significant, which prevents the shell from deviating from the spherical configuration. Accordingly, the shell's configuration roughly resembles a spherical shape. The meager elongation of the shell's frontal side originates from the magnetic force. In this regime, the inner droplet stretches vertically at the rupture instant, which corresponds to $T_i > 0$.

Fig. 24 (b) represents the triangular-like regime, taking place when $Bo \leq 4$ and $Ca \geq 0.15$. In this regime, the viscous drag force is intensified due to the higher values of the Ca number, which flattens the rear side of the shell. Similar to the previous regime, as the magnetic force is inconsequential, the horizontal elongation of the shell is restricted. Consequently, the shell eventually reaches the triangular configuration. Furthermore, in this regime, the inner droplet generally stretches along the *y*-axis.

Revealed in Fig. 24 (c), the egg-like-type 1 regime is observed merely three times for the lowest studied Ca number (0.05) and $2 \leq Bo_m \leq 4$. By virtue of the low Ca number, the interfacial tension force strives to maintain the outer droplet in a spherical configuration. However, the magnetic force elongates both sides of the shell, implying that the shell approaches an ellipsoid shape. Moreover, since the viscous force is negligible relative to the interfacial tension force, the vertical elongating of the inner droplet is suppressed. Hence, intriguingly, the inner droplet does not touch the shell at the upper and lower sides at the breakup moment; instead, solely, the tip of the inner droplet contacts the shell, ending up with an egg-like configuration. In this regime, $T_o$ and $T_i$ are negative, referring to the horizontal elongation.

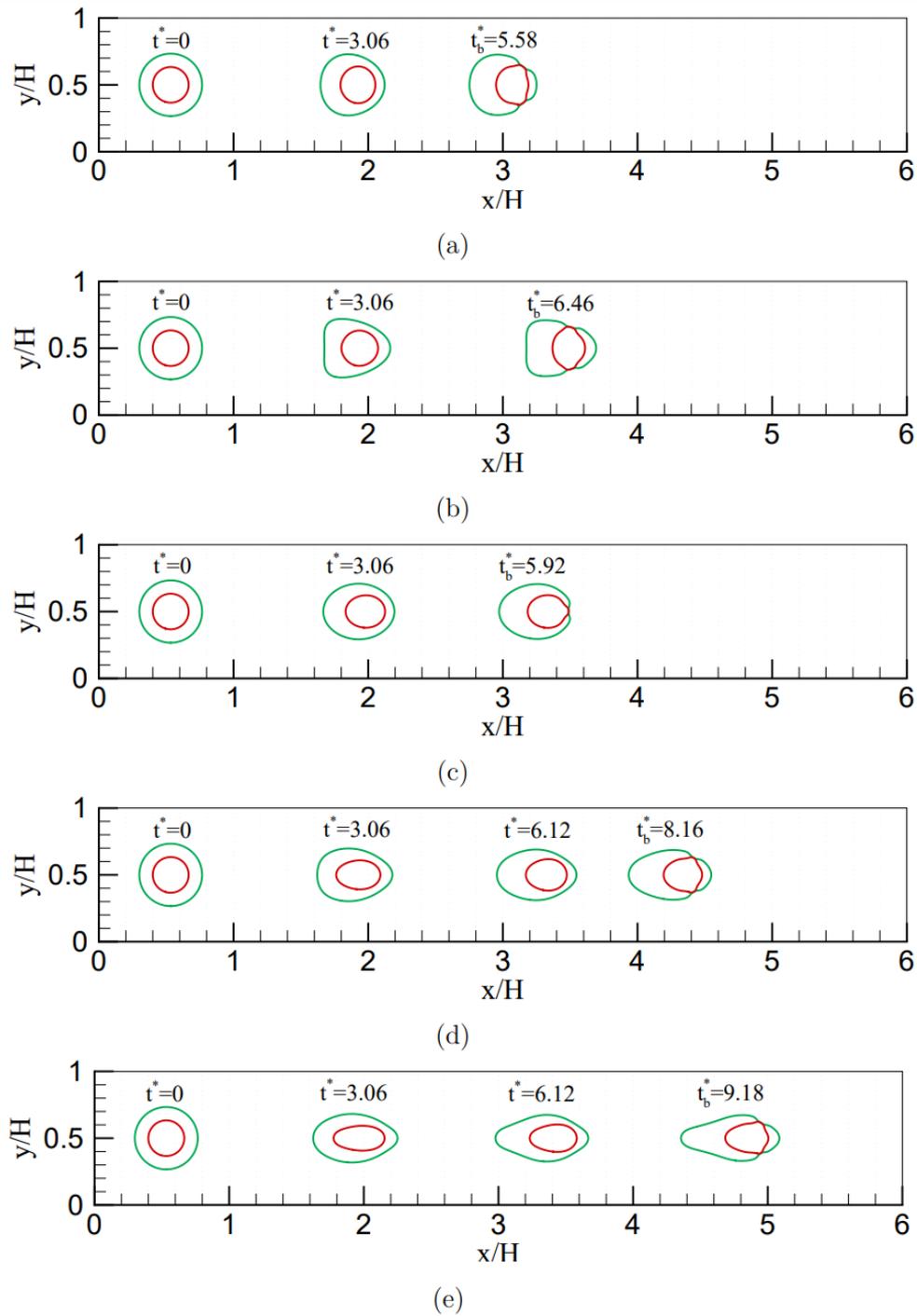

Fig. 24 A typical example of the distinct regimes occurred during the movement of an FCD in a microchannel in the simultaneous presence of the pressure-driven flow and UEMF. The shell is presumed to be ferrofluid and the UEMF is applied along $\alpha = 0°$; (a) semi-spherical regime (Bo = 0, Ca = 0.05), (b) triangular-like regime (Bo = 1, Ca = 0.25), (c) egg-like-type 1 regime (Bo = 2, Ca = 0.05), (d) egg-like-type 2 regime (Bo = 4, Ca = 0.20), and (e) tadpole-like regime (Bo = 7, Ca = 0.25).

Fig. 24 (d) exemplifies the egg-like-type 2 regime, taking place at the moderate $Bo_m$ numbers ($3 \leq Bo_m \leq 5$) for Ca $\geq 0.10$. The morphology of inner and outer droplets resembles the previous regime; by contrast, in this regime, as the Ca number grows, the viscous force instigates the vertical elongation of the inner droplet, causing it to touch the shell at the upper and lower sides instead of a point.

By further augmenting the $Bo_m$ number, the tadpole-like regime arises, as demonstrated in Fig. 24 (e). Observed at high $Bo_m$ numbers, the tadpole-like regime embraces the notable horizontal elongation of the shell induced by the magnetic force. Actually, the shell's rear side becomes longer and narrower, ending up approaching a tail-like structure. On the other hand, the frontal side of the shell's interface remains wider at the breakup moment. Among the five distinguished regimes, the core and shell's deformation is the greatest in the tadpole-like regime. Identical to the previous regime, the core touches the outer droplet at the lower and upper sides when the breakup occurs.

The regime map of the FCD traversing in the microchannel is given in Fig. 25 for a wide range of the $Bo_m$ number (0 – 8) and several Ca numbers ranging from 0.05 to 0.30. The semi-spherical regime is restricted to the region where magnetic and viscous forces are small compared to the interfacial tension force so that the FCD structure does not deviate remarkably from its initial spherical shape. The triangular-like regime emerges when $Ca \leq 0.30$ and $Bo \leq 4$. As the $Bo_m$ number rises, one should boost the Ca number so that this regime happens. That is because the viscous force should be so considerable that it can compete with magnetic force and deform the shell's rear side into a flat shape. As discussed earlier, the egg-like-type 1 regime only arises for the lowest Ca number surveyed and low to moderate $Bo_m$ numbers. By augmenting the Ca number, the transition from the egg-like-type 1 regime to the egg-like-type 2 regime takes place. Ultimately, as the $Bo_m$ number grows, the tadpole-like regime is observed, whose characteristics were delineated previously.

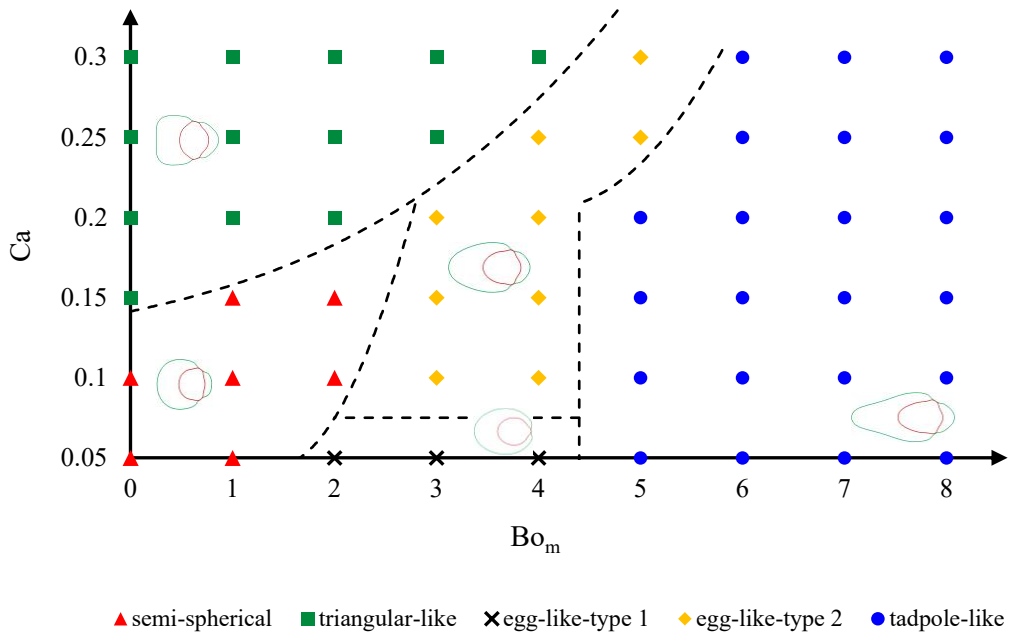

Fig. 25 The regime map of an FCD moving in a microchannel under the concurrent effects of the pressure-driven flow and a UEMF in Bo - Ca parameters. The shell is presumed to be ferrofluid, and the UEMF is applied along α = 0°.

Eventually, Fig. 26 illustrates the variations of the dimensionless breakup length of FCD ($L_b^*$) versus the Ca and $Bo_m$ numbers, when the shell is ferrofluid and $\alpha = 0°$, in which each color corresponds to a specific regime. This diagram unveils how altering the Ca and $Bo_m$ numbers may result in a dramatic increase in breakup length, which alludes to a delay in the rupture incidence. To be pragmatic in a circumstance where one requires to retard the breakup until x/H = 4, in Fig. 26, $L_b^* < 4$ is represented by rectangular cuboids, while $L_b^* \geq 4$ is indicated by cylinders. For $Bo_m < 3$, altering the Capillary number trivially raises the breakup length. Furthermore, when the Ca number is low, augmenting the $Bo_m$ number insignificantly raises the dimensionless breakup length. On the other hand, as the Ca number grows, the impact of the magnetic Bond number on retarding of the rupture process becomes more pronounced. Quantitatively speaking, when Ca = 0.20, augmenting the magnetic Bond number from 1 to 2, results in a 12.0% increment in the breakup length. Moreover, interestingly, in the case of the semi-spherical, triangular-like, and egg-like-type 1 regimes, the dimensionless rupture length is below 4 for all amounts of the Ca and $Bo_m$ numbers, alluding to a prompt rupture. Conversely, for the egg-like-type 2 regime, the rupture length is higher than 4, independent of the $Bo_m$ and Ca values. Also, concerning the tadpole-like regime, except for two cases where Ca = 0.05 and $Bo_m < 7$, the dimensionless rupture length is greater than 4. Consequently, to increase the breakup length until x/H = 4 for a genuine application, one should select the $Bo_m$ and Ca numbers such that the egg-like-type 2 and tadpole-like regimes are obtained.

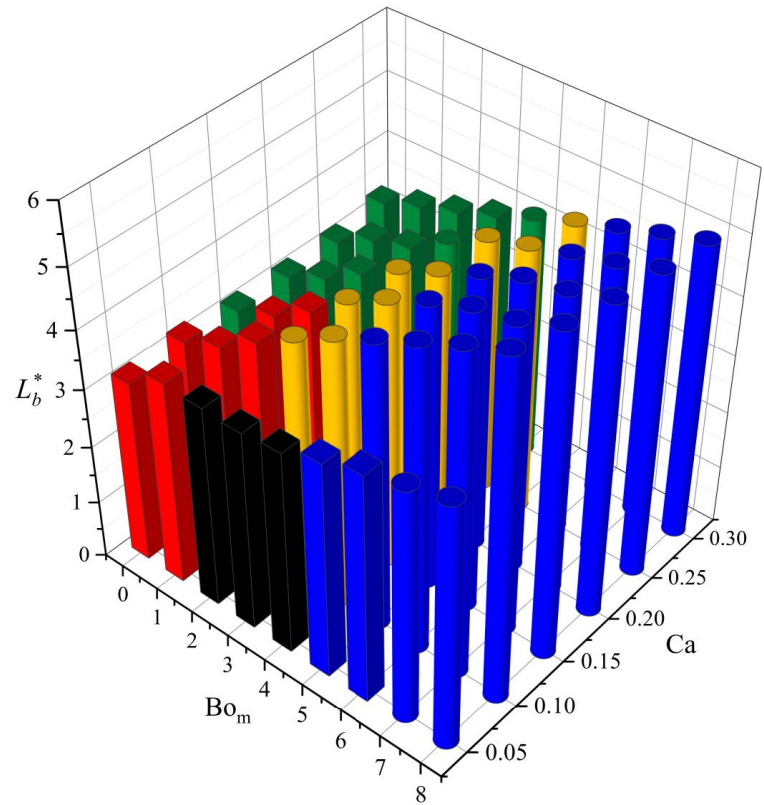

Fig. 26 The dimensionless breakup length ($L_b^*$) with respect to the Ca and $Bo_m$ numbers for five distinct regimes observed during the traversing of an FCD in a microchannel under the impacts of the pressure-driven flow and UEMF. The shell is supposed to be ferrofluid and $\alpha = 0°$. Cylinders represent $L_b^* > 4$, whereas rectangular cuboids demonstrate $L_b^* < 4$.

## 4   Conclusion

For the first time, the hydrodynamics and morphology of a single-core ferrofluid compound droplet (FCD) migrating in a channel flow under a uniform external magnetic field (UEMF) were scrutinized exhaustively. The chief objective of this study is to determine the dominant parameters such that the breakup phenomenon is retarded. The finite-difference method was coupled with the 2D multiple-relaxation time lattice Boltzmann approach to systematically assess the ternary fluid system and magnetic field. The influences of (1) magnetic Bond number ($Bo_m$), (2) Capillary number (Ca), (3) density ratio, (4) viscosity ratio, (5) radius ratio, and (6) surface tension coefficients were investigated on (1) outer droplet deformation ($T_o$), (2) inner droplet deformation ($T_i$), (3) dimensionless breakup time ($t_b^*$), (4) dimensionless breakup length ($L_b^*$), and (5) the breakup height with respect to the channel centerline ($y_b^*$) by presuming either the inner droplet (core) or the outer droplet (shell) is ferrofluid when the UEMF is enforced along $\alpha = 0°$ and $\alpha = 90°$ relative to the fluid flow. Succinctly, the following overriding conclusions may be drawn:

- Regardless of the magnetic field angle, when taken as ferrofluid, the inner droplet undergoes a considerable deformation, while the shell deformation is restricted, gradually confining the region between the core and shell and resulting in a swift breakup. Moreover, as the core is supposed to be ferrofluid, the magnetic force is not brought to bear on the shell. However, interestingly, $T_o$ increases with $\text{Bo}_m$ since the magnetic force elongates the core horizontally, which in turn compresses the fluid inside the shell, boosting the shell deformation.
- Selecting the shell as ferrofluid and imposing the UEMF at $\alpha = 90°$ dramatically precipitates the rupture. Conversely, when applied at $\alpha = 0°$, the magnetic field may retard the breakup process. Hence, the rest of the simulations were conducted at $\alpha = 0°$ by supposing the shell to be ferrofluid.
- When the shell is ferrofluid, the pressure inside the core is augmented, possibly by virtue of the magnetic force direction pulling the core inward.
- Curtailing the Ca number accentuates the interfacial tension force striving to hold the compound droplet in a spherical configuration. Thus, when Ca is low, applying a UEMF brings about an ellipsoid configuration as the magnetic force stretches the frontal and rear sides of the shell simultaneously. In contrast, as the Ca number rises, the viscous force flattens the shell's rear side. Under this condition, enforcing a gentle magnetic force generates a triangular-like shell.
- Though expediting the rupture process, increasing the density ratio (reducing the shell density) negligibly affects $T_i$ and $T_o$.
- Raising the viscosity ratio undermines the drag force applied to the core. Accordingly, the core can transport inside the shell more promptly, accelerating the breakup.
- As the core radius lessens, the region between the core and shell increases, allowing the flow development inside the shell, which in turn enlarges the velocity magnitude in the shell. The velocity increment intensifies the drag force applied to the inner droplet, slowing down its movement and drastically retarding the rupture.

- While reducing the inner droplet deformation, increasing the surface tension coefficient between the core and shell augments the shell deformation. Additionally, by altering the surface tension coefficient such that the complete engulfing-type 1 condition is attained, the rupture length grows by 29.6% compared to a circumstance under which all surface tension coefficients are equal, corresponding to a striking delay in the breakup.

- By varying the $Bo_m$ number from 0 to 8 and the Ca number from 0.05 to 0.30, five distinct breakup regimes were observed: (1) semi-spherical, (2) triangular-like, (3) egg-like-type 1, (4) egg-like-type 2, and (5) tadpole-like, whose regime map was constructed in $Bo_m$ – Ca parameters.

- Surveying the FCD breakup length in various regimes, one may ascertain that by changing the Ca and $Bo_m$ numbers so that the egg-like-type 2 and tadpole-like regimes are accomplished, the breakup length rises, which postpones the rupture process accordingly.

Having an immediate application in drug delivery in which retarding the breakup incidence is favorable, the scholarship endeavors to bridge the gap between FCD hydrodynamics and microfluidics. However, the hydrodynamics and morphology of an FCD under the simultaneous influence of the fluid flow and UEMF is a convoluted topic, entailing more thorough investigations to elucidate all the ins and outs of this paramount issue.

# 5 References

[1] F. Heidari, S.M. Jafari, A.M. Ziaiifar, N. Malekjani, Stability and release mechanisms of double emulsions loaded with bioactive compounds; a critical review, Advances in Colloid and Interface Science, 299 (2022) 102567, doi: https://doi.org/10.1016/j.cis.2021.102567.

[2] J. Wu, C. Gao, D. Sun, L. Yang, B. Ye, T. Wang, P. Zhou, Thermally mediated double emulsion droplets formation in a six-way junction microfluidic device, Colloids and Surfaces A: Physicochemical and Engineering Aspects, 661 (2023) 130961, doi: https://doi.org/10.1016/j.colsurfa.2023.130961.

[3] X. Huang, M. Saadat, M.A. Bijarchi, M.B. Shafii, Ferrofluid double emulsion generation and manipulation under magnetic fields, Chemical Engineering Science, 270 (2023) 118519, doi: https://doi.org/10.1016/j.ces.2023.118519.

[4] F. Peng, Z. Wang, Y. Fan, Q. Yang, J. Chen, Study on the interfacial dynamics of free oscillatory deformation and breakup of single-core compound droplet, Physics of Fluids, 34 (2022), doi: https://doi.org/10.1063/5.0087738.

[5] A. Kumar, R. Kaur, V. Kumar, S. Kumar, R. Gehlot, P. Aggarwal, New insights into water-in-oil-in-water (W/O/W) double emulsions: Properties, fabrication, instability mechanism, and food applications, Trends in Food Science & Technology, 128 (2022) 22-37, doi: https://doi.org/10.1016/j.tifs.2022.07.016.

[6] P. Zhang, L. Xu, H. Chen, A.R. Abate, Flow cytometric printing of double emulsions into open droplet arrays, Lab on a Chip, 23 (2023) 2371-2377, doi: https://doi.org/10.1039/D3LC00151B.

[7] A.L. Guilherme, I.R. Siqueira, L.H.P. Cunha, R.L. Thompson, T.F. Oliveira, Ferrofluid droplets in planar extensional flows: Droplet shape and magnetization reveal novel rheological signatures of ferrofluid emulsions, Physical Review Fluids, 8 (2023) 063601, doi: https://doi.org/10.1103/PhysRevFluids.8.063601.

[8] W. Liu, J.M. Park, Numerical study on the engulfing behavior between immiscible droplets in a confined shear flow, Chemical Engineering Science, 266 (2023) 118265, doi: https://doi.org/10.1016/j.ces.2022.118265.

[9] Z. Ma, S. Zhang, B. Wang, Q. Liu, X. Chen, Deformation characteristics of compound droplets with different morphologies during transport in a microchannel, Physics of Fluids, 35 (2023), doi: https://doi.org/10.1063/5.0146560.

[10] M.P. Borthakur, G. Biswas, D. Bandyopadhyay, Dynamics of deformation and pinch-off of a migrating compound droplet in a tube, Physical Review E, 97 (2018) 043112, doi: https://doi.org/10.1103/PhysRevE.97.043112.

[11] H.V. Vu, T.V. Vu, B.D. Pham, H.D. Nguyen, V.T. Nguyen, H.T. Phan, C.T. Nguyen, Deformation of a compound droplet in a wavy constricted channel, Journal of Mechanical Science and Technology, 37 (2023) 191-202, doi: https://doi.org/10.1007/s12206-022-1220-5.

[12] X. Qu, Y. Wang, Dynamics of concentric and eccentric compound droplets suspended in extensional flows, Physics of Fluids, 24 (2012), doi: https://doi.org/10.1063/1.4770294.

[13] S. Santra, D.P. Panigrahi, S. Das, S. Chakraborty, Shape evolution of compound droplet in combined presence of electric field and extensional flow, Physical Review Fluids, 5 (2020) 063602, doi: https://doi.org/10.1103/PhysRevFluids.5.063602.

[14] G. Hao, L. Li, W. Gao, X. Liu, Y. Chen, Electric-field-controlled deformation and spheroidization of compound droplet in an extensional flow, International Journal of Multiphase Flow, 168 (2023) 104559, doi: https://doi.org/10.1016/j.ijmultiphaseflow.2023.104559.

[15] Y. Chen, X. Liu, M. Shi, Hydrodynamics of double emulsion droplet in shear flow, Applied Physics Letters, 102 (2013), doi: https://doi.org/10.1063/1.4789865.

[16] Z.Y. Luo, L. He, B.F. Bai, Deformation of spherical compound capsules in simple shear flow, Journal of Fluid Mechanics, 775 (2015) 77-104, doi: https://doi.org/10.1017/jfm.2015.286.

[17] H. Hua, J. Shin, J. Kim, Dynamics of a compound droplet in shear flow, International Journal of Heat and Fluid Flow, 50 (2014) 63-71, doi: https://doi.org/10.1016/j.ijheatfluidflow.2014.05.007.

[18] H. Liu, Y. Lu, S. Li, Y. Yu, K.C. Sahu, Deformation and breakup of a compound droplet in three-dimensional oscillatory shear flow, International Journal of Multiphase Flow, 134 (2021) 103472, doi: https://doi.org/10.1016/j.ijmultiphaseflow.2020.103472.

[19] W. Wei, F. Chen, Y. Qiu, L. Zhang, J. Gao, T. Wu, P. Wang, M. Zhang, Q. Zhu, Co-encapsulation of collagen peptide and astaxanthin in WG/OG/W double emulsions-filled alginate hydrogel beads: Fabrication, characterization and digestion behaviors, Journal of Colloid and Interface Science, 651 (2023) 159-171, doi: https://doi.org/10.1016/j.jcis.2023.07.201.

[20] Z. Che, Y.F. Yap, T. Wang, Flow structure of compound droplets moving in microchannels, Physics of Fluids, 30 (2018), doi: https://doi.org/10.1063/1.5008908.

[21] H. Chen, J. Li, H.C. Shum, H.A. Stone, D.A. Weitz, Breakup of double emulsions in constrictions, Soft Matter, 7 (2011) 2345-2347, doi: https://doi.org/10.1039/C0SM01100B.

[22] J. Li, H. Chen, H.A. Stone, Breakup of Double Emulsion Droplets in a Tapered Nozzle, Langmuir, 27 (2011) 4324-4327, doi: https://doi.org/10.1021/la200473h.

[23] N.X. Ho, T.V. Vu, Numerical simulation of the deformation and breakup of a two-core compound droplet in an axisymmetric T-junction channel, International Journal of Heat and Fluid Flow, 86 (2020) 108702, doi: https://doi.org/10.1016/j.ijheatfluidflow.2020.108702.

[24] A. Sattari, N. Tasnim, P. Hanafizadeh, M. Hoorfar, Motion and deformation of migrating compound droplets in shear-thinning fluids in a microcapillary tube, Physics of Fluids, 33 (2021), doi: https://doi.org/10.1063/5.0045994.

[25] C.T. Nguyen, H.V. Vu, T.V. Vu, T.V. Truong, N.X. Ho, B.D. Pham, H.D. Nguyen, V.T. Nguyen, Numerical analysis of deformation and breakup of a compound droplet in microchannels, European Journal of Mechanics - B/Fluids, 88 (2021) 135-147, doi: https://doi.org/10.1016/j.euromechflu.2021.03.005.

[26] V. Thammanna Gurumurthy, S. Pushpavanam, Hydrodynamics of a compound drop in plane Poiseuille flow, Physics of Fluids, 32 (2020), doi: https://doi.org/10.1063/5.0009401.

[27] J. Philip, Magnetic nanofluids (Ferrofluids): Recent advances, applications, challenges, and future directions, Advances in Colloid and Interface Science, 311 (2023) 102810, doi: https://doi.org/10.1016/j.cis.2022.102810.

[28] M.A. Obeid, C.A. Ogah, C.O. Ogah, O.S. Ajala, M.R. Aldea, A.I. Gray, J.I. Igoli, V.A. Ferro, Formulation and evaluation of nanosized hippadine-loaded niosome: Extraction and isolation, physicochemical properties, and in vitro cytotoxicity against human ovarian and skin cancer cell lines, Journal of Drug Delivery Science and Technology, 87 (2023) 104766, doi: https://doi.org/10.1016/j.jddst.2023.104766.

[29] J.D. Benther, B. Wilson, P.A. Petrini, P. Lappas, G. Rosengarten, Ferrofluid droplet impingement cooling of modified surfaces under the influence of a magnetic field, International Journal of Heat and Mass Transfer, 215 (2023) 124370, doi: https://doi.org/10.1016/j.ijheatmasstransfer.2023.124370.

[30] M. Kole, S. Khandekar, Engineering applications of ferrofluids: A review, Journal of Magnetism and Magnetic Materials, 537 (2021) 168222, doi: https://doi.org/10.1016/j.jmmm.2021.168222.

[31] L. Wu, J. Qian, X. Liu, S. Wu, C. Yu, X. Liu, Numerical Modelling for the Droplets Formation in Microfluidics - A Review, Microgravity Science and Technology, 35 (2023) 26, doi: https://doi.org/10.1007/s12217-023-10053-0.

[32] M.A. Bijarchi, A. Favakeh, E. Sedighi, M.B. Shafii, Ferrofluid droplet manipulation using an adjustable alternating magnetic field, Sensors and Actuators A: Physical, 301 (2020) 111753, doi: https://doi.org/10.1016/j.sna.2019.111753.

[33] M.A. Bijarchi, A. Favakeh, S. Alborzi, M.B. Shafii, Experimental investigation of on-demand ferrofluid droplet generation in microfluidics using a Pulse-Width Modulation magnetic field with proposed correlation, Sensors and Actuators B: Chemical, 329 (2021) 129274, doi: https://doi.org/10.1016/j.snb.2020.129274.

[34] M. Aboutalebi, M.A. Bijarchi, M.B. Shafii, S. Kazemzadeh Hannani, Numerical investigation on splitting of ferrofluid microdroplets in T-junctions using an asymmetric magnetic field with proposed correlation, Journal of Magnetism and Magnetic Materials, 447 (2018) 139-149, doi: https://doi.org/10.1016/j.jmmm.2017.09.053.

[35] R.K. Shah, S. Khandekar, On-demand augmentation in heat transfer of Taylor bubble flows using ferrofluids, Applied Thermal Engineering, 205 (2022) 118058, doi: https://doi.org/10.1016/j.applthermaleng.2022.118058.

[36] M.R. Hassan, C. Wang, Magnetic field induced ferrofluid droplet breakup in a simple shear flow at a low Reynolds number, Physics of Fluids, 31 (2019), doi: https://doi.org/10.1063/1.5124134.

[37] S.-T. Zhang, X.-D. Niu, Q.-P. Li, A. Khan, Y. Hu, D.-C. Li, A numerical investigation on the deformation of ferrofluid droplets, Physics of Fluids, 35 (2023), doi: https://doi.org/10.1063/5.0131884.

[38] R.F. Abdo, V.G. Abicalil, L.H.P. Cunha, T.F. Oliveira, On the rheology and magnetization of dilute magnetic emulsions under small amplitude oscillatory shear, Journal of Fluid Mechanics, 955 (2023) A3, doi: https://doi.org/10.1017/jfm.2022.1019.

[39] M. Majidi, M.A. Bijarchi, A.G. Arani, M.H. Rahimian, M.B. Shafii, Magnetic field-induced control of a compound ferrofluid droplet deformation and breakup in shear flow using a hybrid lattice Boltzmann-finite difference method, International Journal of Multiphase Flow, 146 (2022) 103846, doi: https://doi.org/10.1016/j.ijmultiphaseflow.2021.103846.

[40] K. Wang, Y.-C. Xia, Z.-Y. Li, A phase-field Lattice Boltzmann method for liquid-vapor phase change problems based on conservative Allen-Cahn equation and adaptive treegrid, Computers & Fluids, 264 (2023) 105973, doi: https://doi.org/10.1016/j.compfluid.2023.105973.

[41] H.L. Wang, Z.H. Chai, B.C. Shi, H. Liang, Comparative study of the lattice Boltzmann models for Allen-Cahn and Cahn-Hilliard equations, Physical Review E, 94 (2016) 033304, doi: 10.1103/PhysRevE.94.033304.

[42] A. Kumar, Isotropic finite-differences, Journal of Computational Physics, 201 (2004) 109-118, doi: https://doi.org/10.1016/j.jcp.2004.05.005.

[43] Y.Q. Zu, S. He, Phase-field-based lattice Boltzmann model for incompressible binary fluid systems with density and viscosity contrasts, Physical Review E, 87 (2013) 043301, doi: https://doi.org/10.1103/PhysRevE.87.043301.

[44] J. Liu, S.-H. Tan, Y.F. Yap, M.Y. Ng, N.-T. Nguyen, Numerical and experimental investigations of the formation process of ferrofluid droplets, Microfluidics and Nanofluidics, 11 (2011) 177-187, doi: https://doi.org/10.1007/s10404-011-0784-7.

[45] T. Krüger, H. Kusumaatmaja, A. Kuzmin, O. Shardt, G. Silva, E.M. Viggen, The lattice Boltzmann method, Springer International Publishing, 10 (2017) 4-15, doi: https://doi.org/10.1007/978-3-319-44649-3.

[46] P. Poureslami, M. Siavashi, H. Moghimi, M. Hosseini, Pore-scale convection-conduction heat transfer and fluid flow in open-cell metal foams: A three-dimensional multiple-relaxation time lattice Boltzmann (MRT-LBM) solution, International Communications in Heat and Mass Transfer, 126 (2021) 105465, doi: https://doi.org/10.1016/j.icheatmasstransfer.2021.105465.

[47] A. Fakhari, T. Mitchell, C. Leonardi, D. Bolster, Improved locality of the phase-field lattice-Boltzmann model for immiscible fluids at high density ratios, Physical Review E, 96 (2017) 053301, doi: https://doi.org/10.1103/PhysRevE.96.053301.

[48] P. Lallemand, L.-S. Luo, Theory of the lattice Boltzmann method: Dispersion, dissipation, isotropy, Galilean invariance, and stability, Physical Review E, 61 (2000) 6546-6562, doi: 10.1103/PhysRevE.61.6546.

[49] A. Fakhari, T. Lee, Multiple-relaxation-time lattice Boltzmann method for immiscible fluids at high Reynolds numbers, Physical Review E, 87 (2013) 023304, doi: https://doi.org/10.1103/PhysRevE.87.023304.

[50] S. Chen, D. Martínez, R. Mei, On boundary conditions in lattice Boltzmann methods, Physics of Fluids, 8 (1996) 2527-2536, doi: https://doi.org/10.1063/1.869035.

[51] G.R. McNamara, G. Zanetti, Use of the Boltzmann Equation to Simulate Lattice-Gas Automata, Physical Review Letters, 61 (1988) 2332-2335, doi: https://doi.org/10.1103/PhysRevLett.61.2332.

[52] R. Haghani Hassan Abadi, M.H. Rahimian, A. Fakhari, Conservative phase-field lattice-Boltzmann model for ternary fluids, Journal of Computational Physics, 374 (2018) 668-691, doi: https://doi.org/10.1016/j.jcp.2018.07.045.

[53] C. Bracco, C. Giannelli, A. Reali, M. Torre, R. Vázquez, Adaptive isogeometric phase-field modeling of the Cahn–Hilliard equation: Suitably graded hierarchical refinement and coarsening on multi-patch geometries, Computer Methods in Applied Mechanics and Engineering, (2023) 116355, doi: https://doi.org/10.1016/j.cma.2023.116355.

[54] F. Ren, B. Song, M.C. Sukop, H. Hu, Improved lattice Boltzmann modeling of binary flow based on the conservative Allen-Cahn equation, Physical Review E, 94 (2016) 023311, doi: https://doi.org/10.1103/PhysRevE.94.023311.

[55] Y. Li, D. Jeong, H. Kim, C. Lee, J. Kim, Comparison study on the different dynamics between the Allen–Cahn and the Cahn–Hilliard equations, Computers & Mathematics with Applications, 77 (2019) 311-322, doi: https://doi.org/10.1016/j.camwa.2018.09.034.

[56] Y. Sun, C. Beckermann, Sharp interface tracking using the phase-field equation, Journal of Computational Physics, 220 (2007) 626-653, doi: https://doi.org/10.1016/j.jcp.2006.05.025.

[57] P.-H. Chiu, Y.-T. Lin, A conservative phase field method for solving incompressible two-phase flows, Journal of Computational Physics, 230 (2011) 185-204, doi: https://doi.org/10.1016/j.jcp.2010.09.021.

[58] F. Boyer, C. Lapuerta, Study of a three component Cahn-Hilliard flow model, ESAIM: Mathematical Modelling and Numerical Analysis, 40 (2006) 653-687, doi: https://doi.org/10.1051/m2an:2006028.

[59] R. Haghani Hassan Abadi, A. Fakhari, M.H. Rahimian, Numerical simulation of three-component multiphase flows at high density and viscosity ratios using lattice Boltzmann methods, Physical Review E, 97 (2018) 033312, doi: https://doi.org/10.1103/PhysRevE.97.033312.

[60] F. Boyer, C. Lapuerta, S. Minjeaud, B. Piar, M. Quintard, Cahn–Hilliard/Navier–Stokes Model for the Simulation of Three-Phase Flows, Transport in Porous Media, 82 (2010) 463-483, doi: https://doi.org/10.1007/s11242-009-9408-z.

[61] T.R. Mitchell, M. Majidi, M.H. Rahimian, C.R. Leonardi, Computational modeling of three-dimensional thermocapillary flow of recalcitrant bubbles using a coupled lattice Boltzmann-finite difference method, Physics of Fluids, 33 (2021), doi: https://doi.org/10.1063/5.0038171.

[62] N. Wang, C. Semprebon, H. Liu, C. Zhang, H. Kusumaatmaja, Modelling double emulsion formation in planar flow-focusing microchannels, Journal of Fluid Mechanics, 895 (2020) A22, doi: https://doi.org/10.1017/jfm.2020.299.

[63] T.-V. Vu, T.V. Vu, C.T. Nguyen, P.H. Pham, Deformation and breakup of a double-core compound droplet in an axisymmetric channel, International Journal of Heat and Mass Transfer, 135 (2019) 796-810, doi: https://doi.org/10.1016/j.ijheatmasstransfer.2019.02.032.

[64] S. Santra, S. Das, S. Chakraborty, Electric field-induced pinch-off of a compound droplet in Poiseuille flow, Physics of Fluids, 31 (2019), doi: https://doi.org/10.1063/1.5094948.

[65] M.A. Bijarchi, M. Yaghoobi, A. Favakeh, M.B. Shafii, On-demand ferrofluid droplet formation with non-linear magnetic permeability in the presence of high non-uniform magnetic fields, Scientific Reports, 12 (2022) 10868, doi: https://doi.org/10.1038/s41598-022-14624-w.

[66] T.-V. Vu, T.V. Vu, D.T. Bui, Numerical study of deformation and breakup of a multi-core compound droplet in simple shear flow, International Journal of Heat and Mass Transfer, 131 (2019) 1083-1094, doi: https://doi.org/10.1016/j.ijheatmasstransfer.2018.11.131.

[67] J. Guzowski, P.M. Korczyk, S. Jakiela, P. Garstecki, The structure and stability of multiple micro-droplets, Soft Matter, 8 (2012) 7269-7278, doi: https://doi.org/10.1039/C2SM25838B.